\documentclass[%
 reprint,
 amsmath,amssymb,
 aps,
]{revtex4-2}

\usepackage[normalem]{ulem}

\usepackage{graphicx}
\usepackage{dcolumn}
\usepackage{bm}

\usepackage{graphics}

\usepackage{physics}

\usepackage{float}

\usepackage{euscript}

\newcommand{\fh}{\boldsymbol{\hat{f}}_\lambda}
\newcommand{\fhd}{\boldsymbol{\hat{f}}^{\dagger}_{\lambda}}
\newcommand{\fhp}{\boldsymbol{\hat{f}}_{\lambda'}}
\newcommand{\fhdp}{\boldsymbol{\hat{f}}^{\dagger}_{\lambda'}}

\newcommand{\Hh}{\hat{H}} 

\newcommand{\ah}{\hat{a}}
\newcommand{\ahdot}{\dot{\hat{a}}}
\newcommand{\ahd}{\hat{a}^{\dagger}}
\newcommand{\ahddot}{\dot{\hat{a}}^{\dagger}}

\newcommand{\bh}{\hat{b}}
\newcommand{\bhd}{\hat{b}^{\dagger}}

\newcommand{\Dh}{\hat{D}}
\newcommand{\Dhd}{\hat{D}^{\dagger}}

\newcommand{\hh}{\hat{h}}
\newcommand{\hhd}{\hat{h}^{\dagger}}

\newcommand{\sh}{\hat{\sigma}} 
\newcommand{\shd}{\hat{\sigma}^{\dagger}} 

\newcommand{\rb}{\vb{r}}

\newcommand{\Gbl}{\vb{G}_{\lambda}}
\newcommand{\Gbld}{{\vb{G}_{\lambda}^{*}}^\mathrm{T}}

\newcommand{\Gbldp}{{\vb{G}_{\lambda'}^{*}}^\mathrm{T}}
\newcommand{\Gb}{\vb{G}}
\newcommand{\kpar}{k^{\parallel}}

\newcommand{\kparsq}{k^{\parallel^2}}

\newcommand{\Eb}{\vb{E}}

\newcommand{\Pc}{\mathcal{P}}
\newcommand{\Qc}{\mathcal{Q}}
\newcommand{\Lc}{\mathcal{L}}
\newcommand{\Dc}{\mathcal{D}}

\begin{document}

\preprint{APS/123-QED}

\title{Nested Open Quantum Systems Approach to Photonic Bose--Einstein Condensation}

\author{Stefan Yoshi Buhmann}
 
\affiliation{%
 Institut f\"ur Physik, Universit\"at Kassel, Heinrich-Plett Stra{\ss}e 40, 34132 Kassel, Germany
}%

\author{Andris Erglis}
\thanks{andris.erglis@physik.uni-freiburg.de}%

\affiliation{
 Physikalisches Institut, Albert-Ludwigs-Universit\"at Freiburg, Hermann-Herder-Stra{\ss}e 3, 79104 Freiburg, Germany
}%

\date{\today}

\begin{abstract}
The photonic Bose--Einstein condensate is a recently observed collective ground state of a coupled light-matter system. We describe this quantum state based on macroscopic quantum electrodynamics in dispersing and absorbing environments. To model the coupled photon--dye dynamics, we derive a master equation using a nested open quantum systems approach yielding all parameters essential to describe the condensation process. This approach allows us to describe photon condensates of arbitrary shapes because all geometry-dependent decay constants can be expressed in terms of the Green’s tensor. In particular, we obtain the cavity mode absorption and emission rates of the dye molecules.
\end{abstract}

\maketitle

\section{Introduction}
Photonic Bose--Einstein condensation is a phenomenon where photons reach thermal equilibrium with a well-defined effective temperature and macroscopically occupy the lowest energy state possible in the system.  It is analogous to an atomic Bose--Einstein condensate (BEC), but the conditions for photons to achieve this state are different.
The first BECs were observed with atoms \cite{anderson1995, bradley1995evidence, davis1995bose,bradley1997bose}, and since then experiments have demonstrated BECs of magnons, polaritons and excitons \cite{demokritov2006bose, serga2014bose, balili2007bose, carusotto2013quantum, kasprzak2006bose, deng2002condensation}. 

For a long time, it was assumed that photons cannot form a BEC because of their non-interacting nature and because they disappear in the cavity walls when decreasing the temperature, as described in the black-body radiation model. It turns out that photon condensate is possible if one uses a dye-filled microcavity with highly reflective mirrors \cite{klaers2010bose,klaers2011bose, marelic2015experimental, nyman2018bose}. The dye allows for the photons to reach thermal equilibrium through multiple absorption/emission cycles. The thermalization time must be much faster than the rate at which photons are lost from cavity modes due to spontaneous or cavity decay. When the photons reach thermal equilibrium, their effective temperature is equal to the temperature of the dye whose absorption/emission spectrum must obey the Kennard-Stepanov relation \cite{mccumber1964einstein}.

The mirrors provide a trapping potential that endows the photons with an effective mass and prevents them from escaping the cavity before reaching thermalization. Condensation can be achieved at room temperature when a critical number of photons in the system is reached.

Recent reports demonstrate photon condensation with 68 photons \cite{dung2017variable} and even as few as 7 photons \cite{walker2018driven}. Other work has demonstrated the condensation of photons inside a one-dimensional fiber cavity \cite{weill2019bose}. Here, the thermalization of photons was achieved by many interaction cycles of photons and Er/Yb. One of the recent reports demonstrates the coupling of condensate through tunneling by exploiting two minima in a mirror potential where symmetric and anti-symmetric eigenstates of the condensate have been formed \cite{kurtscheid2019thermally}.

The BEC is a promising candidate for many applications, e.g., atomic and photonic lasers \cite{wiseman1997defining, andrews1997observation, durfee1998experimental, bloch1999atom, rajan2016photon, muller2019general}, atomic interferometry \cite{cronin2009optics, altin2011optically} and in quantum information processing \cite{nyman2018bose}. Despite being a quantum phenomenon, condensation has also been theoretically predicted and observed for classical light, allowing applications in imaging \cite{aschieri2011condensation, sun2012observation}.

On the theory side, one of the first works \cite{muller1986bose} has shown that a photon BEC can be achieved if the grand canonical ensemble can be applied for photons. 
Some of the theoretical work has 
assumed that photons are already in the thermal equilibrium \cite{klaers2012statistical,sob2012hierarchical},
while other work has 
focused on the non-equilibrium dynamics describing the condensation from laser theory by considering two-level atoms interaction with photons \cite{chiocchetta2014quantum}. The latter does not take into account the rovibrational coupling of dye molecules which is an important ingredient for the thermalization of photons. One of the recent papers 
demonstrates that photon condensation can occur in three dimensions with thermalization mechanisms other than dye \cite{muller2019general}. 

So far, the most in-depth work is the microscopic model developed by Keeling and Kirton \cite{kirton2013nonequilibrium,kirton2015thermalization}, which provides the non-equilibrium dynamics of the photons and can describe fluctuations and correlations of the condensate. It allows for predicting the photon condensation threshold by considering multiple parameters important in the experiment. Two of them, namely, absorption/emission rates of the dye molecules, are derived using an open quantum systems description while others, namely, spontaneous and cavity decay rate, and incoherent pumping rate are included phenomenologically. The theory has been further developed to describe the spatial profile of the photon BEC with respect to the spot size of the laser pump and polarization dynamics in the condensate \cite{keeling2016spatial, moodie2017polarization}. The extended version of this model has also been used to study the transition between BEC, multimode condensation and lasing \cite{hesten2018decondensation}.

In this paper, we develop a first-principle theory of photonic Bose-Einstein condensation by combining open quantum systems and macroscopic quantum electrodynamics (QED) \cite{breuer2002theory, buhmann2013dispersion}. There are three advantages to our approach. First, we derive all dynamic equations of condensate from first principles. Second, this approach allows us to calculate all the necessary parameters
which is particularly useful for planning new experiments.
Third, it is formally possible to calculate the condensate dynamics for arbitrary geometries 
since we use Green's tensor formalism. It could be applied, for example, for cascaded mirrors with two dimples \cite{kurtscheid2019thermally}, periodic structures \cite{dung2017variable} or even arbitrarily-shaped potentials \cite{kurtscheid2020realizing}.

The theoretical model consists of multiple systems and baths interacting with each other. In Section \ref{Section:Nested_Systems_General_Description} we give a short introduction to nested open quantum systems. In Section \ref{Section:ConstructingHamiltonian} we start the development of the theory by constructing the Hamiltonian of the photon--molecule interaction inside the cavity. Next, we exploit the concept of nested open quantum systems: in step 1 (see Section \ref{Section:Step1}), we derive the master equation in Lindblad form for the cavity and molecular decay and pumping constants. Then, in step 2 (see Section \ref{Section:Step2}), we separate the remaining system Hamiltonian into the new system, bath, and interaction parts and derive the master equation for absorption/emission rates, which are now influenced by the previously derived rates from step 1. Along with the derivation, we demonstrate that spontaneous molecular decay and absorption/emission rates are proportional to Green's tensor and the dipole moment of the dye molecule.

\section{Nested Open Quantum Systems}\label{Section:Nested_Systems_General_Description}

\begin{figure}
    \includegraphics[scale=0.3]{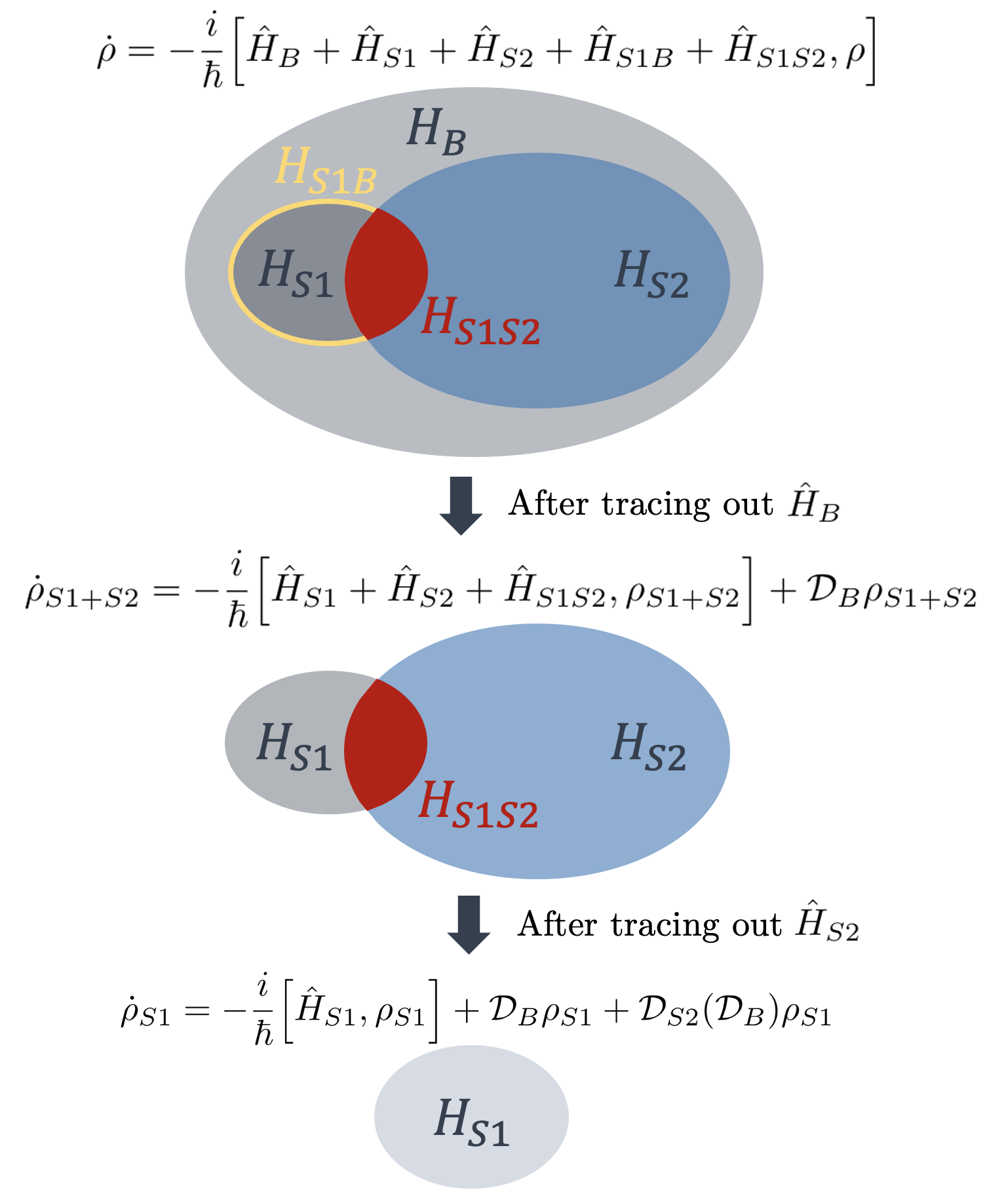}
    \caption{Schematic representation of a nested open quantum systems approach. We have two systems described by $\Hh_{S1}$ and $\Hh_{S2}$ embedded into an environment described by $H_{B}$. In step 1 we trace out $H_B$ to obtain the master equation for density matrix $\rho_{S1+S2}$ where $\mathcal{D}_B$ is the Markov approximated Lindblad dissipator. In step 2 we treat $H_{S2}$ as an environment and trace it out to obtain master equation for $\rho_{S1}$ with an additional Lindblad dissipator $\Dc_{S2}(\Dc_B)$ which is now a function of (influenced by) $\Dc_B$. }
    \label{fig:system_bath_general}
\end{figure}

\begin{figure*}
    \includegraphics[scale=0.2]{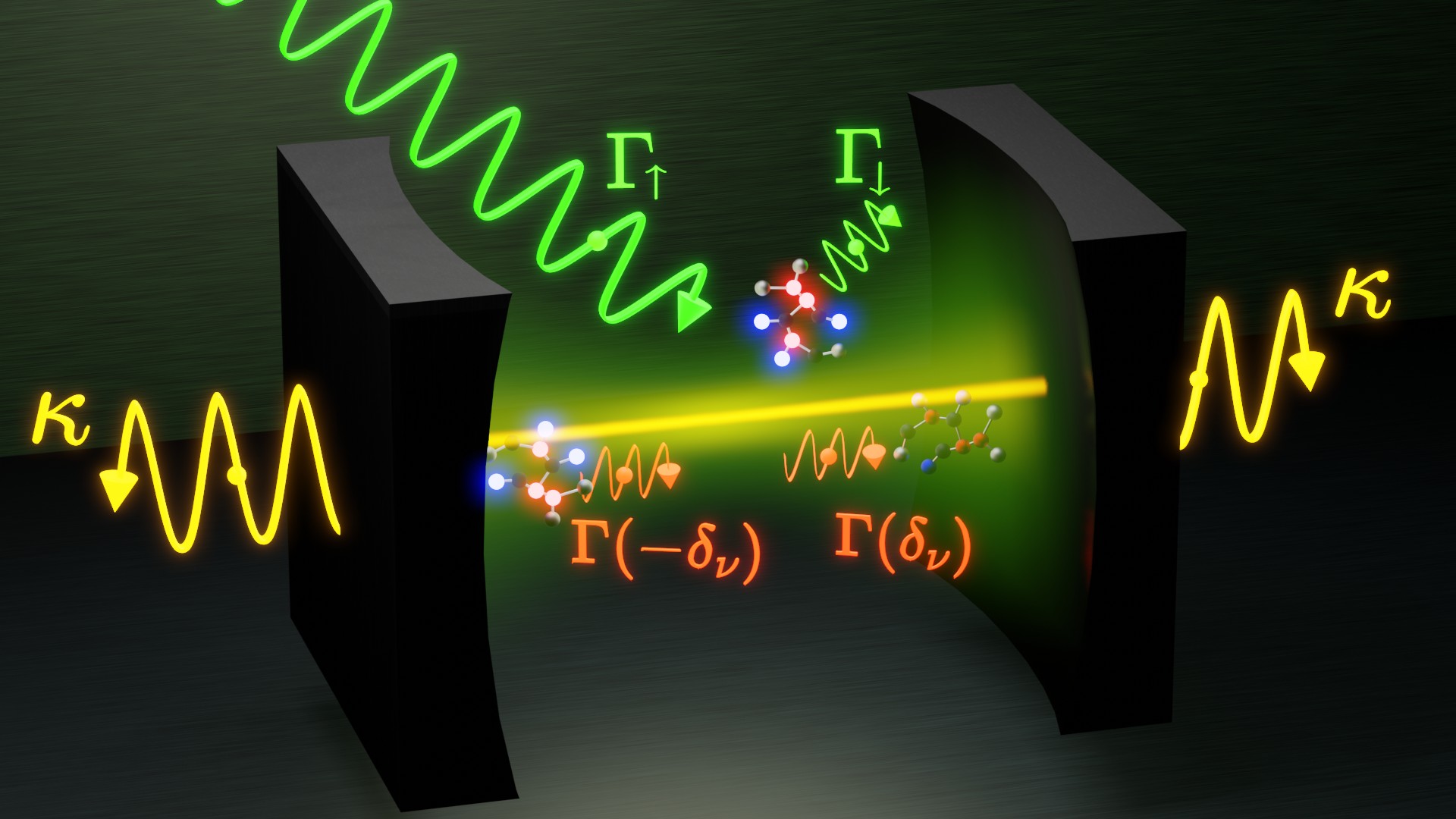}
    \caption{Photon BEC setup. Dye molecules inside a cavity are pumped with a laser at a rate $\Gamma_{\uparrow}$. The molecules can emit photons into a cavity mode $\nu$ with a rate $\Gamma(-\delta_{\nu})$ and absorb them with a rate $\Gamma(\delta_{\nu})$ . Excitations are lost from the cavity by spontaneous emission from the molecule with a rate $\Gamma_{\downarrow}$ or by a cavity decay with a rate $\kappa$. }
    \label{fig:cavity}
\end{figure*}

Figure \ref{fig:system_bath_general} shows the concept of a nested open quantum systems. The goal of this approach is to describe the interaction of a system (here S$_1$) with two baths (B and S$_2$) where the second bath itself is subject to dissipative dynamics due to the influence of the first bath. This is achieved in a two-step process. Firstly, we have two Hamiltonians $\Hh_{S1}$ and $\Hh_{S2}$ describing two systems $S1$ and $S2$ interacting with each other via interaction term $\Hh_{S1S2}$. They both are immersed in a common bath described by $\Hh_B$. For simplicity, we assume that only $S1$ is directly interacting with the bath via term $\Hh_{S1B}$.

Starting from the unitary evolution of the total density matrix $\rho$ (top of figure), we trace out $\Hh_B$. This leads to the dissipative dynamics of the density matrix $\rho_{S1+S2}$, where $\Dc_B$ is the dissipator in Lindblad form (middle of the figure). Next, we trace out $\Hh_{S2}$ as the new bath. This results in an additional dissipative term $\Dc_{S2}(\Dc_B)$ which is a function of $\Dc_B$. Thus, the nested systems approach allows us to capture the influence of $\Dc_{B}$ onto $\Dc_{S2}$.

\section{Hamiltonian}\label{Section:ConstructingHamiltonian}

Figure \ref{fig:cavity} shows a schematic illustration of a model to describe a photon BEC. It consists of a cavity made of two highly reflective mirrors where in between there are dye molecules. Photons from the laser enter the cavity via the dye molecules by being absorbed (with a pumping rate $\Gamma_{\uparrow}$) and emitted into the cavity mode $\nu$ [with a rate $\Gamma(-\delta_{\nu})$]. Molecules can also absorb the photons from the cavity mode [with a rate $\Gamma(\delta_{\nu})$]. There are two ways photons are lost from the cavity: they may leak from the cavity mirrors (with a rate $\kappa$) or spontaneously decay into a non-cavity mode (with a rate $\Gamma_{\downarrow}$). A photon BEC is formed once a critical number of photons (or the pumping rate $\Gamma_{\uparrow}$) is exceeded inside the cavity, which is illustrated in Fig. \ref{fig:cavity} as a bright yellow line around the optical axis.

The starting point is a  well-known macroscopic QED description of molecules, photons and their interaction \cite{buhmann2013dispersion2}. Here we employ a model of identical molecules $i$ as two-level systems where each level is dressed by rovibrational states. Electronic levels of the molecule are represented by Pauli matrices $\sh_i$ with the electronic molecular transition frequency being $\omega_{10}$. The rovibrational (phonon) modes are described by harmonic oscillators with mode operators $\bh_i$ and $\bhd_i$ and the transition frequency between modes $\Omega$. Here, $\omega_{10}$ and $\Omega$ are assumed to be the same for all molecules. Note that $\Omega\ll\omega_{10}$. We assume that the molecule is described by only one rovibrational mode. In principle, it is possible to include multiple electronic and phononic modes that would give rise to multiple peaks in the absorption/emission spectrum of the molecule. However, the simplified model is sufficient to describe the photon BEC dynamics. The two-level system and rovibrational states couple with each other with coupling strength given by Huang--Rhys parameter $S$. The Hamiltonian for a total of $N$ molecules then reads:
\begin{equation}\label{Eq:MoleculeHamiltonian}
  \Hh_M=\sum_{i=1}^N \left[\frac{\hbar}{2}\omega_{10}\sh_i^z+\hbar\Omega \bhd_i\bh_i  +  \hbar \Omega \sqrt{S} \sh_i^z(\bh_i+\bhd_i)\right].  
\end{equation}

To describe cavity-assisted photons, we quantize the field in media and obtain the fundamental field operators $\fh(\rb, \omega)$ and $\fhd(\rb, \omega)$ at position $\rb$ and frequency $\omega$. They are related to the polarization and magnetization of the medium \cite{buhmann2013dispersion} and obey the commutation relation $\comm{\fh(\rb',\omega')}{\fhdp(\rb,\omega)}=\delta_{\lambda \lambda'}\delta(\omega-\omega')\vb{\delta}(\rb-\rb')$. The total photon field can then be expressed as the sum of electric and magnetic excitations e, m and integral of $\fh(\rb, \omega)$ and $\fhd(\rb, \omega)$ over the entire space in position and frequency:
\begin{equation}\label{Eq:FieldHamiltonian}
\Hh_F=\sum\limits_{\lambda = \mathrm{e, m}} \int \dd^3r \int\limits_0^\infty \dd\omega \, \hbar \omega \, \fhd(\rb, \omega) \cdot \fh(\rb, \omega).
\end{equation}

Interactions between the molecules and photons take place when a photon gets absorbed (emitted) from a ground (excited) state of a two-level system. The molecule--field interaction is then taken in dipole approximation to be:
\begin{equation}\label{Eq:InteractionHamiltonian}
    \Hh_{MF}=-\sum_i(\vb{d}_{01}\sh_i+\vb{d}_{10}\shd_i)\cdot \hat{\Eb}(\rb_i),
\end{equation}
where $\vb{d}_{10}$ is the dipole moment and $\hat{\Eb}(\rb_i)$ is the electric field operator at molecule's position $\rb_i$. The electric field can be expressed in terms of the Green's tensor $\Gbl$ and operators $\fh$ \cite{buhmann2013dispersion}:
\begin{multline}
    \hat{\Eb}(\rb_i)=\sum\limits_{\lambda = e, m} \int \dd^3r \int\limits_0^\infty \dd\omega \, \bigg\{  \Gbl(\rb_i, \rb, \omega) \cdot \fh(\rb, \omega) \\
    + 
    \fhd(\rb, \omega) \cdot \Gbld(\rb_i, \rb, \omega)  \bigg\},
\end{multline}
where $\Gb_e$ and $\Gb_m$ are defined as:
\begin{equation}\label{Eq:GreensTensorDefinition}
\begin{aligned}
    &\Gb_e(\rb, \rb', \omega)=\mathrm{i}\frac{\omega^2}{c^2}\sqrt{\frac{\hbar}{\pi \varepsilon_0}\Im\varepsilon(\rb', \omega)}\Gb(\rb, \rb', \omega),\\
    &\Gb_m(\rb, \rb', \omega)\!=\!\mathrm{i}\frac{\omega}{c}\sqrt{\frac{\hbar}{\pi \varepsilon_0}\frac{\Im\mu(\rb', \omega)}{\abs{\Im\mu(\rb', \omega)}^2}}[\bm{\nabla}'\!\cross\!\Gb(\rb', \rb, \omega)]^{\mathrm{T}},
\end{aligned}
\end{equation}
with $\varepsilon(\rb, \omega)$ and $\mu(\rb, \omega)$ being the electric permittivity and magnetic permeability of the medium, $\varepsilon_0$ being vacuum permittivity and $c$ being the speed of light.

\begin{figure}
    \centering
    \includegraphics[scale=0.35]{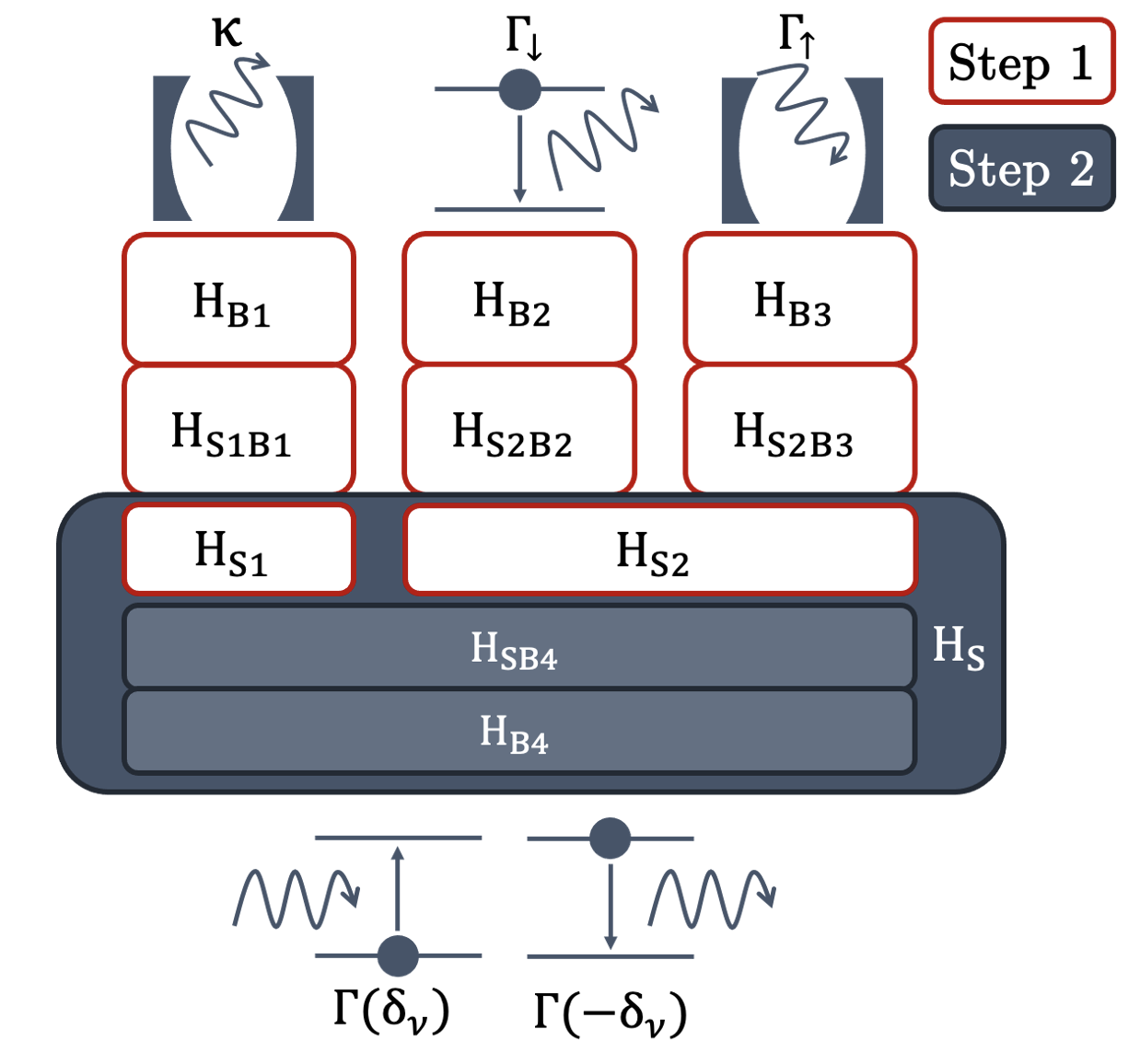}
    \caption{Schematic representation of a nested open quantum systems approach for our model. In step 1 we derive the rates $\kappa, \Gamma_{\downarrow}$ and $\Gamma_{\uparrow}$ from baths B1, B2 and B3, and their interaction with systems S1 and S2. There are additional terms $H_{B4}$ and $H_{SB4}$ which we treat as part of the system in step 1. In step 2, they take on the roles of bath and system-bath interaction and from their interaction with $H_{S1}$ and $H_{S2}$ we derive $\Gamma(\pm\delta_{\nu})$ .}
    \label{fig:system_bath}
\end{figure}

As we have now stated the Hamiltonian describing photons, molecules and their interactions, we will identify different parts of the Hamiltonian responsible for different dissipation processes.
As described in Introduction and Section \ref{Section:Nested_Systems_General_Description}, we will use a nested open quantum systems approach. Fig. \ref{fig:system_bath} shows a schematic diagram of a nested open quantum systems for our model. In step 1, there are baths responsible for cavity and spontaneous decay and laser pumping (see Fig. \ref{fig:system_bath}). The system Hamiltonians $\Hh_{S1}$ and $\Hh_{S2}$ are coupled to the bath Hamiltonians $\Hh_{B1}$,  $\Hh_{B2}$ and $\Hh_{B3}$ through the respective interaction Hamiltonians. Once we derive the master equation for the corresponding system, in step 2, we separate the remaining system Hamiltonian again into the system, bath, and interaction Hamiltonians. Then we trace out the bath and obtain the emission and absorption rates for the photons in the cavity modes $\Gamma(\pm\delta_{\nu})$.

The main reason why we exploit this approach is that in general, both $\Gamma(\pm\delta_{\nu})$ are influenced (broadened) by incoherent processes in the cavity, in this case, $\kappa, \Gamma_{\downarrow}$ and $\Gamma_{\uparrow}$. Exploiting the nested approach allows us to capture this influence. 

To derive the master equation with dissipative constants of interest, the first step is to perform a separation of the total photon field into different fields which we will treat as baths responsible for dissipative processes. 
Thus, we separate $\Hh_F$ into the laser field (responsible for $\Gamma_{\uparrow}$, see Appendix \ref{appendix:LaserDriving}), resonant cavity modes, the remaining field responsible for spontaneous decay ($\Gamma_{\downarrow}$), and cavity leakage ($\kappa$). Hereby, we will assign each Hamiltonian a system (S), bath (B), or interaction (SB) nomenclature because it will be useful later when deriving the master equation.

Within the laser source region $V_\mathrm{L}$ there is a coherent field and everywhere else there are vacuum fluctuations. Thus, we can separate the total field state as follows \cite{fuchs2018casimir}:
\begin{equation}\label{Eq:LaserState}
    \ket{\psi}_{\mathrm{F}}=\underset{\rb\in V_{\mathrm{L}}}{|\{\fh(\rb, \omega\}}\rangle\otimes\underset{\rb\notin V_{\mathrm{L}}}{\ket{\{0\}}}.
\end{equation}

If we act with an annihilation operator on this state we obtain the number instead of an operator at the laser source:
\begin{equation}
    \fh(\rb, \omega)\ket{\psi}_{\mathrm{F}}=
    \begin{cases}
    \boldsymbol{f}_{\lambda}(\rb, \omega)\ket{\psi}_{\mathrm{F}} &\rb\in V_{\mathrm{L}},\\
    0 &\rb\notin V_{\mathrm{L}}.
    \end{cases}
\end{equation}
We can then separate Eq. \eqref{Eq:FieldHamiltonian} as follows:
\begin{multline}\label{Eq:FieldHamiltonian2}
\Hh_F=\Hh_F(\rb\notin V_{\mathrm{L}})+\Hh_F(\rb\in V_{\mathrm{L}})\\=\sum\limits_{\lambda = \mathrm{e, m}} \int_{\rb\notin V_{\mathrm{L}}} \dd^3r \int\limits_0^\infty \dd\omega \, \hbar \omega \, \fhd(\rb, \omega) \cdot \fh(\rb, \omega)\\+\sum\limits_{\lambda = \mathrm{e, m}} \int_{\rb\in V_{\mathrm{L}}}  \dd^3r \int\limits_0^\infty \dd\omega \, \hbar \omega \, \fhd(\rb, \omega) \cdot \fh(\rb, \omega).
 \end{multline}

Next, we assume that the laser field is independent of the position $\rb$. Because the laser field is pumping molecules, it works as an amplifier in the open quantum systems context, thus, we redefine the second part of Eq. \eqref{Eq:FieldHamiltonian2} in terms of inverted oscillator operators $\hh(\omega'')$ and define it to be the bath 3 Hamiltonian \cite{gardiner2004quantum}:
    \begin{equation}\label{Eq:GlauberHamiltonian}
    \Hh_F(\rb\in V_{\mathrm{L}}) \equiv \Hh_{B3} = -\int \dd \omega'' \, \hbar \omega'' \, \hh(\omega'')\hhd(\omega'').
    \end{equation}
Note, that the inverted oscillator operator $\hh(\omega'')$ and its Hermitian conjugate is exactly opposite to the photon field operator $\fh(\rb, \omega)$ and it obeys the commutation relation:
\begin{equation}\label{Eq:InverseCommutation}
    \comm{\hh(\omega)}{\hhd(\omega')}=-\delta(\omega-\omega').
\end{equation}
The average values are given the opposite of usual photon occupation number:
\begin{align}\label{Eq:InverseOccupation}
    &\langle \hhd(\omega)\hh(\omega)\rangle=(N(\omega)+1)\delta(\omega-\omega'),\\
    &\langle \hh(\omega)\hhd(\omega)\rangle=N(\omega)\delta(\omega-\omega'),
\end{align}
where $N(\omega)$ is the Bose-Einstein distribution with a negative laser temperature $T_{laser}<0$:
\begin{equation}\label{Eq:InvertedPhotonNumber}
    N(\omega)=\frac{1}{e^{-\hbar\omega/k_B T_{laser}}-1}.
\end{equation}

The next step is to separate the field of the Hamiltonian $\Hh_F(\rb\notin V_{\mathrm{L}})$ into that of the cavity modes and that of the remaining field. The cavity modes are described by operators $\ah_{m \zeta}$ and $\ahd_{m \zeta}$, where the cavity resonance has a Lorentzian shape. They can destroy (create) a photon of a cavity mode with labels $m$, $\zeta$. As an example for the planar cavity, $m$ determines the number of standing-wave modes along the cavity axis, while $\zeta$ is the mode by the transversal wavenumber $\kpar$, which has an expression $\kparsq=k^2-(m\pi/d)^2$ where $d$ is the length of the cavity and $k$ is the total wavenumber. The annihilation operator reads \cite{oppermann2018quantum}: 
\begin{equation}\label{Eq1:BosonOperatorsCavity}
    \ah_{m\zeta}=\sqrt{\frac{\gamma_{m\zeta}}{2\pi}}\int\limits_{-\infty}^{\infty} \dd \omega \, \frac{\ah(\omega,\zeta)}{\omega-\omega_{m\zeta}+\mathrm{i}\gamma_{m\zeta}/2},
\end{equation}
where $\ah(\omega)\equiv\ah(\rb_i,\omega)$ at molecule's position $\rb_i$ is defined through Green's tensor (see Eq. \eqref{Eq:GreensTensorDefinition}) $\Gbl$ \cite{buhmann2013dispersion2}:
\begin{multline}
       \ah(\omega, \zeta) = -\frac{1}{\hbar g(\rb_i, \omega_{m\zeta}, \zeta)}\\ \times \sum\limits_{\lambda = e, m} \int \dd^3r' \int\limits_0^\infty \dd\omega \, \vb{d}_{10} \cdot \Gbl(\rb, \rb', \omega) \cdot \fh(\rb', \omega) ,
\end{multline}
with the interaction strength between photons and molecules being
\begin{multline}
    g^2(\rb, \omega_{m\zeta}, \zeta) \\= \frac{\mu_0}{\pi \hbar} \omega_{m\zeta}^2 \vb{d}_{10} \cdot \Im \Gb (\rb, \rb, \omega_{m\zeta}, \zeta) \cdot\vb{d}_{01} = \frac{\Omega_R^2}{2\pi \gamma_{m\zeta}},
\end{multline}
where $\Omega_R=\sqrt{ \frac{2}{ \hbar}\mu_0 \gamma_{m\zeta} \omega_{m\zeta}^2  \vb{d}_{10} \cdot \Im \Gb (\rb, \rb, \omega_{m\zeta}, \zeta) \cdot\vb{d}_{01}}$ is the Rabi frequency, $\gamma_{m\zeta}$ is the width of the resonance at the resonant frequency $\omega_{m\zeta}$, and $\mu_0$ is the magnetic permeability. Note that $$\Gb(\rb, \rb', \omega)=\int \dd\zeta \Gb(\rb, \rb', \omega, \zeta).$$

For notation convenience, we subsume the two mode labels within a multi-index $\nu$, i.e., $m\zeta\equiv \nu$. We separate the first part of Eq. \eqref{Eq:FieldHamiltonian2} into the Hamiltonian of the resonant cavity modes and a remaining field bath by adding and subtracting $\Hh_{S1}=\sum\hbar \omega_{\nu} \ahd_{\nu}\ah_{\nu}$:
 \begin{equation}\label{Eq:RemainingFieldHamiltonian}
    \Hh_F(\rb\notin V_{\mathrm{L}})=\Hh_{RF}+\Hh_{S1},
\end{equation}
where we the remaining field (RF) reads
\begin{multline}\label{Eq:RFHamiltonian}
    \Hh_{RF}=\sum\limits_{\lambda = \mathrm{e, m}} \int_{\rb\notin V_{\mathrm{L}}} \dd^3r \int\limits_0^\infty \dd\omega \, \hbar \omega \, \fhd(\rb, \omega) \cdot \fh(\rb, \omega)\\-\sum_{\nu=1}^{\infty}\hbar \omega_{\nu} \ahd_{\nu}\ah_{\nu}.
\end{multline}
$\Hh_{RF}$ is responsible for the spontaneous decay of the molecules, thus we label it as the bath 2 Hamiltonian $\Hh_{RF}\equiv \Hh_{B2}$. 

To account for the cavity leakage, we need to add another bath Hamiltonian. It turns out that the remaining field Hamiltonian can account not only for the spontaneous decay but also for the leakage of the cavity modes. But, for the interaction Hamiltonian with cavity modes to be well defined, it has to be recast into a different form, which we define as the bath 1 Hamiltonian:
\begin{equation}\label{Eq:Bath1Hamiltonian}
   \Hh_{B1} = \Hh_{RF}=\int\limits_0^\infty \dd\omega' \, \hbar \omega' \, \ahd(\omega') \ah(\omega').
\end{equation}

Now that we have separated the total field into multiple baths of interest, we do a similar procedure for the molecule--field interaction term \eqref{Eq:InteractionHamiltonian}. 
For the setup of interest, the laser frequency is near-resonant $\omega\approx\omega_{10}$, thus, we can perform the rotating wave approximation (RWA) and the interaction Hamiltonian \eqref{Eq:InteractionHamiltonian} reads:
\begin{multline}
    \Hh_{MF}=- \sum_i^N \sum\limits_{\lambda = e, m}\int \dd^3r \int\limits_0^\infty \dd\omega \, \\ \times 
    \bigg\{ \vb{d}_{10} \cdot \Gbl(\rb_i, \rb, \omega, \zeta) \cdot \fh(\rb, \omega) \shd_i\\
     + 
    \fhd(\rb, \omega) \cdot \Gbld(\rb_i, \rb, \omega, \zeta) \cdot \vb{d}_{01}  \sh_i \bigg\}.
\end{multline}
Now we proceed in a similar manner as we did for the photon field by separating the interaction term. First, we split $\Hh_{MF}$ into:
\begin{equation}
    \Hh_{MF}=\Hh_{MF}(\rb\notin V_{\mathrm{L}})+\Hh_{MF}(\rb \in V_{\mathrm{L}}).
\end{equation}
The interaction part $\Hh_{MF}(\rb \in V_{\mathrm{L}})$ for the laser is semi-classical as can be obtained if $\hat{\vb{E}}(\rb_i)$ is applied to the state in Eq. \eqref{Eq:LaserState} \cite{fuchs2018casimir}. The explicit expression is given by Eq. \eqref{Eq:LaserDrivingHamiltonian}. But to employ the open quantum systems context, similarly as in Eq. \eqref{Eq:GlauberHamiltonian}, we redefine $\Hh_{MF}(\rb \in V_{\mathrm{L}})$ to be as a quantized interaction between system 2 and bath 3:
\begin{multline}
    \Hh_{MF}(\rb \in V_{\mathrm{L}})\equiv\Hh_{S2B3} \\= \sum_i\int\limits_0^\infty \dd \omega'' \, \hbar \mu(\omega'')\left(\hh(\omega'')\shd_i+\hhd(\omega'')\sh_i\right),
\end{multline}
with the coupling strength $\mu(\omega'')$. In Appendix \ref{appendix:LaserDriving} we demonstrate how the pumping term $\Gamma_{\uparrow}$ is related to the laser parameters. The system 2 Hamiltonian from Eq. \eqref{Eq:MoleculeHamiltonian} is identified as:
\begin{equation}
  \Hh_{S2}=\sum_{i=1}^N\frac{\hbar}{2}\omega_{10}\sh_i^z.
\end{equation}

Treating the interaction term $\Hh_{MF}(\rb\notin V_{\mathrm{L}})$, in the same manner as we did for the field in Eq. \eqref{Eq:RemainingFieldHamiltonian}, we add and subtract the resonant interaction term $\sum_{\nu,i}  \frac{1}{2}\hbar \Omega_R(\rb_i, \omega_{\nu}, \zeta)[\ah_{\nu}\shd_i+\ahd_{\nu}\sh_i]$ which is analogous to Jaynes--Cummings model. Then we define the remaining-field interaction (RI) term as:
\begin{multline}\label{Eq:RemainingInteraction}
    \Hh_{RI}= \\-\sum\limits_{\lambda = e, m} \sum_i^N \int \dd^3r \int\limits_0^\infty \dd\omega \, \bigg\{ \vb{d}_{10} \cdot \Gbl(\rb_i, \rb, \omega) \cdot \fh(\rb, \omega) \shd_i\\
     + 
    \fhd(\rb, \omega) \cdot \Gbld(\rb_i, \rb, \omega) \cdot \vb{d}_{01}  \sh_i \bigg\} 
    \\ -\sum_{\nu,i}  \frac{1}{2} \hbar \Omega_R(\rb_i, \omega_{\nu}, \zeta)[\ah_{\nu}\shd_i+\ahd_{\nu}\sh_i].
\end{multline}

The other resonant interaction term together with the third part of Eq. \eqref{Eq:MoleculeHamiltonian} we define as system Hamiltonian:
\begin{multline}
    \Hh_{SB4} = \sum_{\nu,i}  \frac{1}{2} \hbar \Omega_{\nu}[\ah_{\nu}\shd_i+\ahd_{\nu}\sh_i] \\ +  \sum_i \hbar \Omega \sqrt{S} \sh_i^z(\bh_i+\bhd_i),
\end{multline}
where $\Omega_{\nu}\equiv \Omega_R(\rb_i, \omega_{\nu}, \zeta)$.

The remaining-field interaction Hamiltonian $\Hh_{RI}$ is related to the spontaneous decay, thus we define it to be the interaction Hamiltonian between bath 2 and system 2, $\Hh_{RI}\equiv \Hh_{S2B2}$.

To account for the cavity leakage, we need to add the interaction between the remaining field from Eq. \eqref{Eq:Bath1Hamiltonian} with the cavity modes from the second part of Eq. \eqref{Eq:RFHamiltonian}:
\begin{multline}
    \Hh_{S1B1} = \sum_{\nu}\int\limits_0^\infty \dd\omega' \,  \hbar \lambda(\omega')[\ah(\omega')\ahd_{\nu}+\ahd(\omega')\ah_{\nu}],
\end{multline}
with the coupling strength $\lambda(\omega')$. In Appendix \ref{appendix:CavityDecay} we show that the cavity leakage $\kappa$ is related to the width of the cavity resonance $\gamma_{\nu}$.

Finally, the rovibrational states Hamiltonian from Eq. \eqref{Eq:MoleculeHamiltonian} in the open quantum systems context we define as the bath 4 Hamiltonian: 
\vspace{-1mm}
\begin{equation}
  \Hh_{B4}=\sum_i\hbar\Omega \bhd_i\bh_i.
\end{equation}
We treat it as a system term in step 1 and as a bath term in step 2, respectively.

The complete Hamiltonian describing the model is $\Hh=\Hh_{S1}+\Hh_{S2}+\Hh_{B1}+\Hh_{B2}+\Hh_{B3}+\Hh_{S1B1}+\Hh_{S2B2}+\Hh_{S2B3}+\Hh_{B4}+\Hh_{SB4}$ (see Fig. \ref{fig:system_bath}).

\section{Intermediate Master Equation}\label{Section:Step1}

\subsection{Transforming Into the Interaction Picture}

As we have separated the the system of interest in the respective bath, interaction and system Hamiltonians, we are ready to derive the dynamics of the reduced system of photons and electronic transitions of molecules with all the relevant constants.

The first step is to transform interaction Hamiltonians to the interaction picture which read:
\begin{widetext}
\begin{equation}
    \Hh_{S1B1}(t) = \sum_{\nu}\int\limits_0^\infty \dd\omega' \,  \hbar \lambda(\omega')[\ah(\omega')e^{-\mathrm{i}\omega' t}\ahd_{\nu} e^{\mathrm{i}\omega_{\nu} t}+\ahd(\omega')e^{\mathrm{i}\omega' t}\ah_{\nu} e^{-\mathrm{i}\omega_{\nu} t}],
\end{equation}

\vspace{-3mm}
\begin{eqnarray}\label{Eq:HamiltoniansInteractionPicture}
    &&\Hh_{S2B2}(t)\nonumber\\&& = -\sum\limits_{\lambda = e, m}\sum_i^N \int \dd^3r \int\limits_0^\infty \dd\omega \Big [\, \vb{d}_{10} \cdot \Gbl(\rb_i, \rb, \omega) \cdot \fh(\rb, \omega)e^{-\mathrm{i}\omega t} \shd_i e^{\mathrm{i}\omega_{10} t} 
    + \fhd(\rb, \omega)e^{\mathrm{i}\omega t} \cdot \Gbld(\rb_i, \rb, \omega) \cdot \vb{d}_{01}  \sh_i  e^{-\mathrm{i}\omega_{10} t}
    \nonumber \Big ]\\
    &&-\sum_{\nu,i} \frac{1}{2}  \hbar \Omega_R(\rb_i, \omega_{\nu}, \zeta)\Bigg[\sqrt{\frac{\gamma_{m\zeta}}{2\pi}}\int\limits_{-\infty}^{\infty} \dd \omega \, \frac{\ah(\omega,\zeta)e^{-\mathrm{i}\omega t}}{\omega-\omega_{m\zeta}+\mathrm{i}\gamma_{m\zeta}/2} \shd_i e^{\mathrm{i}\omega_{10} t}+\sqrt{\frac{\gamma_{m\zeta}}{2\pi}}\int\limits_{-\infty}^{\infty} \dd \omega \, \frac{\ahd(\omega,\zeta)e^{\mathrm{i}\omega t}}{\omega-\omega_{m\zeta}-\mathrm{i}\gamma_{m\zeta}/2}  \sh_i e^{-\mathrm{i}\omega_{10} t}\Bigg],
\end{eqnarray}

\vspace{-3mm}
\begin{equation}
      \Hh_{S2B3}(t) = \sum_i\int\limits_0^\infty \dd \omega'' \, \hbar \mu(\omega'')\left[\hh(\omega'')e^{-\mathrm{i}\omega'' t}\shd_i e^{\mathrm{i}\omega_{10} t}+\hhd(\omega'')e^{\mathrm{i}\omega'' t}\sh_i e^{-\mathrm{i}\omega_{10} t}\right].
\end{equation}

\end{widetext}

We have used the following approximations: Firstly, as the rovibrational energies are typically much smaller than the electronic and photon energies,   $\Omega\ll\omega_{\nu}, \omega, \omega', \omega''$, we can neglect interaction picture contribution from $\Hh_{S4}$ and $\Hh_{S5}$. Secondly, we assume that the interaction between atoms and photons is not ultrastrong, $g\ll\omega_{\nu}, \omega, \omega', \omega''$, so that we can also neglect the contribution from $\Hh_{S3}$ as well.

\subsection{Constructing the Master Equation}

We start with the usual Markov-approximated density matrix equation in the interaction picture \cite{breuer2002theory}:

\begin{equation}\label{Eq:MasterEqHamiltonian}
\begin{aligned}
    \dot{\tilde{\rho}}_S(t)&=
    \frac{1}{\hbar^2} \int\limits_0^{\infty} \dd \tau \, 
    \\ \times \bigg\{
    &\tr_{B1}\comm{H_{S1B1}(t)}{\comm{H_{S1B1}(t-\tau)}{\tilde{\rho}_S(t)\rho_{B1}}}
    \\+
    &\tr_{B_{\mathrm{tot}}}\comm{H_{S2B2}(t)}{\comm{H_{S2B2}(t-\tau)}{\tilde{\rho}_S(t)\rho_{B_{\mathrm{tot}}}}}\\
    \\-
    &\tr_{B_\mathrm{r}}\comm{H_{S2B2}(t)}{\comm{H_{S2B2}(t-\tau)}{\tilde{\rho}_S(t) \rho_{B_\mathrm{r}}}}
    \\+
    &\tr_{B3}\comm{H_{S2B3}(t)}{\comm{H_{S2B3}(t-\tau)}{\tilde{\rho}_S(t)\rho_{B3}}}
    \bigg\},
\end{aligned}
\end{equation}
where $\tilde{\rho}_S(t)$ is the system density matrix in the interaction picture.
It has been shown \cite{rivas2010markovian} (Appendix B2) that if the interaction between baths is weak, then for each bath there is one Lindbladian superoperator and we can neglect cross terms between different baths. Thus, we split the double commutator and its trace into three parts for each bath.
Since for bath $B2$ we have separated the resonant photon modes from the total field in the Hamiltonian $H_{B2}$, we need to also account for that when calculating the bath density matrix. Thus, we split the density matrix by writing $\rho_{B2}=\rho_{B_{\mathrm{tot}}}-\rho_{B_\mathrm{r}}$ where $\rho_{B_\mathrm{r}}$ and $\rho_{B_{\mathrm{tot}}}$ accounts for the cavity-photon and total bath density matrices, respectively.

To expand the double commutator in Eq. \eqref{Eq:MasterEqHamiltonian}, we identify the system $A_i(t)$ and bath $B_i(t)$ operators by writing the interaction Hamiltonian in the form $\sum_{\alpha } A_{\alpha}(t)\otimes B_{\alpha}(t)$. In this way $\Hh_{S1B1}$ leads to:

\begin{equation}
\begin{aligned}
& A_1(t) = \sum_{\nu} \ahd_{\nu} e^{\mathrm{i}\omega_{\nu} t},  \\
&A_2(t) =\sum_{\nu} \ah_{\nu} e^{-\mathrm{i}\omega_{\nu} t}, 
\end{aligned}
\end{equation}
\begin{equation}
\begin{aligned}
& B_1(t) = \int \limits_0^{\infty} \dd \omega'\hbar \lambda(\omega')\ah(\omega')e^{-\mathrm{i}\omega't},   \\
&B_2(t) = \int \limits_0^{\infty} \dd \omega' \hbar \lambda(\omega')\ahd(\omega')e^{\mathrm{i}\omega't},
\end{aligned}
\end{equation}
whereas $\Hh_{S2B2}$ implies:
\begin{equation}
\begin{aligned}
& A_3(t) = \sum_i \shd_i e^{\mathrm{i}\omega_{10}t},     \\
&A_4(t) = \sum_i \sh_i e^{-\mathrm{i}\omega_{10}t},   
\end{aligned}
\end{equation}
\begin{equation}\label{Eq:Bath2Operators}
\begin{aligned}
&  B_3(t) = -\sum\limits_{\lambda = e, m} \int \dd^3r \int\limits_0^\infty \dd\omega \, \vb{d}_{10} \cdot \Gbl \cdot \fh  e^{-\mathrm{i}\omega t},     \\
&B_4(t) = -\sum\limits_{\lambda = e, m} \int \dd^3r \int\limits_0^\infty \dd\omega \, \fhd \cdot \Gbld \cdot \vb{d}_{01} e^{\mathrm{i}\omega t}
\end{aligned}
\end{equation}
and 
\begin{equation}
\begin{aligned}
& A_{3'}(t) = \sum_i \shd_i e^{\mathrm{i}\omega_{10}t},    \\
&  A_{4'}(t) = \sum_i \sh_i e^{-\mathrm{i}\omega_{10}t},   
\end{aligned}
\end{equation}

\begin{equation}
\begin{aligned}
\begin{split}
&B_{3'}(t)= -\sum_{\nu} \frac{1}{2}\hbar \Omega_R(\rb_i, \omega_{\nu}, \zeta)\\&\times\sqrt{\frac{\gamma_{m\zeta}}{2\pi}}\int\limits_{-\infty}^{\infty} \dd \omega \, \frac{\ah(\omega,\zeta)e^{-\mathrm{i}\omega t}}{\omega-\omega_{m\zeta}+\mathrm{i}\gamma_{m\zeta}/2} 
\end{split}
\\
\begin{split}
&B_{4'}(t) = -\sum_{\nu} \frac{1}{2}\hbar \Omega_R(\rb_i, \omega_{\nu}, \zeta) \\&\times \sqrt{\frac{\gamma_{m\zeta}}{2\pi}}\int\limits_{-\infty}^{\infty} \dd \omega \, \frac{\ahd(\omega,\zeta)e^{\mathrm{i}\omega t}}{\omega-\omega_{m\zeta}-\mathrm{i}\gamma_{m\zeta}/2} ,
\end{split}
\end{aligned}
\end{equation}
and finally, from $\Hh_{S2B3}$ we deduce:
\begin{equation}
\begin{aligned}
& A_5(t) =\sum_i  \shd_i e^{\mathrm{i}\omega_{10}t},  \\
&A_6(t) = \sum_i   \sh_i  e^{-\mathrm{i}\omega_{10} t},   
\end{aligned}
\end{equation}
\vspace{-2mm}
\begin{equation}
\begin{aligned}
&  B_5(t) = \int\limits_0^\infty \dd \omega'' \, \hbar \mu(\omega'')\hh(\omega'')e^{-\mathrm{i}\omega'' t}, \\
& B_6(t) =\int\limits_0^\infty \dd \omega'' \, \hbar \mu(\omega'') \hhd(\omega'')e^{\mathrm{i}\omega'' t}.
\end{aligned}
\end{equation}
Note that $A_3(t)=A_{3'}(t)=A_5(t)$ and $A_4(t)=A_{4'}(t)=A_6(t)$.
The correlation functions read, for example, for the first bath:
\begin{gather}
    C_{12}(\tau)=\tr_{B1}\{B_1 B_2(-\tau) \rho_{B1}\},\\
    C_{21}(\tau)=\tr_{B1}\{B_2 B_1(-\tau) \rho_{B1}\}.
\end{gather}
Because the baths are independent from each other, all correlations between different baths yield zero: $C_{13}=C_{23}=C_{14}=C_{35}=...=0$. Note that all correlations with the same index yield zero as well, i.e. $C_{11}=C_{22}=...=0$, since the average of the product of the same bath operators vanishes. Transforming back to the Schrödinger picture and by expanding the double commutator in the integral we obtain the master equation

\begin{widetext}
\begin{equation}\label{Eq:BigMasterEquation1_1}
\begin{aligned}
     \Dot{\rho}_S(t) = -\frac{\mathrm{i}}{\hbar} \comm{\Hh_S}{\rho_S(t)}
     - 
     \frac{1}{\hbar^2}\int\limits_0^{\infty} \dd\tau 
     \bigg\{&
     \sum_{\nu} 
     \comm{\ahd_{\nu}}{\ah_{\nu} e^{\mathrm{i}\omega_{\nu} \tau}\rho_S(t)}C_{12}(\tau) 
     + 
     \comm{\rho_S(t)\ah_{\nu} e^{\mathrm{i}\omega_{\nu} \tau}}{\ahd_{\nu}}C_{21}(-\tau)\\ 
     +&  
     \comm{\ah_{\nu}}{\ahd_{\nu} e^{-\mathrm{i}\omega_{\nu} \tau}\rho_S(t)}C_{21}(\tau)
     + 
     \comm{\rho_S(t)\ahd_{\nu} e^{-\mathrm{i}\omega_{\nu} \tau}}{\ah_{\nu}}C_{12}(-\tau)\\[2em] 
     +&
    \sum_i 
     \comm{\shd_i}{\sh_i e^{\mathrm{i}\omega_{10}\tau}\rho_S}[C_{34}(\tau)-C_{3'4'}(\tau)]
     +
     \comm{\rho_S(t)\sh_i e^{\mathrm{i}\omega_{10}\tau}}{\shd_i}[C_{43}(-\tau)-C_{4'3'}(-\tau)]\\ 
     +&  
     \comm{\sh_i}{\shd_i e^{-\mathrm{i}\omega_{10}\tau}\rho_S(t)}[C_{43}(\tau)-C_{4'3'}(\tau)]
     +
     \comm{\rho_S(t)\shd_i e^{-\mathrm{i}\omega_{10}\tau}}{\sh_i}[C_{34}(-\tau)-C_{3'4'}(-\tau)]\\[2em] 
     +&  
     \sum_i
     \comm{\shd_i}{\sh_i e^{\mathrm{i}\omega_{10}\tau}\rho_S(t)}C_{56}(\tau) 
     + 
     \comm{\rho_S(t)\sh_i  e^{\mathrm{i}\omega_{10}\tau}}{  \shd_i}C_{65}(-\tau)\\ 
     +&  
     \comm{ \sh_i}{ \shd_i e^{-\mathrm{i}\omega_{10}\tau}\rho_S(t)}C_{65}(\tau) 
     + 
     \comm{\rho_S(t) \shd_i e^{-\mathrm{i}\omega_{10}\tau}}{ \sh_i}C_{56}(-\tau)
     \bigg\},
\end{aligned}
\end{equation}
where $\rho_S(t)$ is the system density matrix in the Schrödinger picture and $\Hh_S=\Hh_{S1}+\Hh_{S2}+\Hh_{B4}+\Hh_{SB4}$.
\end{widetext}

\subsection{Evaluating Environmental Correlation Functions}

Assuming the thermal excitations to be much smaller than the electronic and photonic cavity excitations in the system \cite{klaers2010bose}, i.e, $\omega, \omega_{\nu}, \omega' \gg k_B T /\hbar $, and $\omega''\gg k_B T_{laser} /\hbar$,  we can neglect the respective thermal occupation numbers, $n(\omega)=n(\omega_{\nu})=n(\omega')=N(\omega'')\approx 0$. 

We present the calculation of the first bath correlation functions in detail, which read:
\begin{subequations}
\begin{eqnarray}
      &&C_{12}(\tau)=\langle B_1 B_2(-\tau) \rangle\approx \int\limits_0^\infty \dd\omega'  \,  \hbar^2 \lambda^2( \omega') e^{-\mathrm{i} \omega' t}, \\
    &&C_{21}(\tau) \approx 0. 
\end{eqnarray}
\end{subequations}
Next, we use the relation 
\begin{equation}\label{Eq:IntegralFourier}
    \int\limits_0^{\infty} d\tau e^{\pm \mathrm{i}\omega\tau} = \pi \delta(\omega) \pm \mathrm{i} \pv \frac{1}{\omega},
\end{equation}
to evaluate time--frequency integrals with $\delta(\omega)$ being the Dirac delta function and $\pv$ being the Cauchy principal value. For example the time integral of the coefficient $C_{12}(\tau)$ is given by
\begin{multline}
    \frac{1}{\hbar^2}\int\limits_0^{\infty} \dd \tau \, C_{12}(\tau) e^{\mathrm{i}\omega_{\nu} \tau} \\=\frac{1}{\hbar^2}\int\limits_0^{\infty} \dd \tau \,\int\limits_0^\infty \dd\omega'  \,  \hbar^2 \lambda^2( \omega') e^{-\mathrm{i} (\omega'-\omega_{\nu}) \tau}\\=\pi\lambda^2(\omega_{\nu})- \mathrm{i} \pv \int\limits_0^{\infty} \frac{\dd\omega'}{\omega'-\omega_{\nu}}    \lambda^2(\omega').
\end{multline}
Finally, we define the cavity decay rate $\kappa$ and the Lamb shift $\Delta_{\kappa}$ as:
\begin{subequations}
\begin{align}
    \frac{\kappa}{2} &= \pi \lambda^2(\omega_{\nu}),\\
    \Delta_{\kappa} &= \pv    \int\limits_0^{\infty} \frac{\dd\omega'}{\omega'-\omega_{\nu}} \lambda^2(\omega').
\end{align}
\end{subequations}
In Appendix \ref{appendix:CavityDecay} we demonstrate that $\kappa$ is equal to the width of the cavity resonance $\gamma_{\nu}$.

The correlation functions from the second bath read (derivation in Appendix  \ref{appendix:GreenTensor}):
\begin{gather}\label{Eq:CorrelationC12}
    C_{34}(\tau) \approx \int\limits_0^\infty \dd\omega \, \frac{\hbar\mu_0}{\pi} \omega^2 \vb{d}_{10} \cdot \Im\Gb(\rb_i, \rb_i, \omega)  \cdot \vb{d}_{01}e^{-\mathrm{i}\omega \tau},\\
    C_{43}(\tau) \approx 0.
\end{gather}
Again, we evaluate time--frequency integral using Eq. \eqref{Eq:IntegralFourier}:
\begin{multline}
     \frac{1}{\hbar^2} \int\limits_0^{\infty} \dd \tau \, C_{34}(\tau) e^{\mathrm{i}\omega_{10}\tau}=\frac{\mu_0}{\hbar} \omega_{10}^2 \vb{d}_{10} \cdot \Im\Gb(\rb_i, \rb_i, \omega_{10})  \cdot \vb{d}_{01} \\-\mathrm{i} \pv \int\limits_0^{\infty} \frac{\dd\omega}{\omega-\omega_{10}}    \frac{\mu_0}{\hbar\pi} \omega^2 \vb{d}_{10} \cdot \Im\Gb(\rb_i, \rb_i, \omega)  \cdot \vb{d}_{01}.
\end{multline}
Similarly as before, we define the spontaneous decay rate of all photon modes $\Gamma^{\mathrm{tot}}_{\downarrow}$ and the shift $\Delta_{\Gamma^{\mathrm{tot}}_{\downarrow}}$ as:
\begin{subequations}
\begin{align}
    \frac{\Gamma^{\mathrm{tot}}_{\downarrow}}{2} &=  \frac{\mu_0}{\hbar} \omega_{10}^2 \vb{d}_{10} \cdot \Im\Gb(\rb_i, \rb_i, \omega_{10})  \cdot \vb{d}_{01}, \\
    \Delta_{\Gamma^{\mathrm{tot}}_{\downarrow}} &=   \pv \int\limits_0^{\infty} \frac{\dd\omega}{\omega-\omega_{10}}    \frac{\mu_0}{\hbar\pi} \omega^2 \vb{d}_{10} \cdot \Im\Gb(\rb_i, \rb_i, \omega)  \cdot \vb{d}_{01}.
 \end{align}
\end{subequations}

For the resonant correlation functions we have a sum of interaction strength over modes $\nu$:
\begin{equation}\label{Eq:CorrelationC3'4'}
    C_{3'4'}(\tau)=\sum_{\nu}\frac{1}{4} \hbar^2 \Omega_R^2 \frac{\gamma_{\nu}}{2\pi}\int\limits_{-\infty}^{\infty} \dd \omega \, \frac{e^{-\mathrm{i}\omega t}}{(\omega-\omega_{\nu})^2+\gamma^2_{\nu}/4}.
\end{equation}
\begin{equation}
        C_{4'3'}(\tau) \approx 0.
\end{equation}
Then we evaluate the time--frequency integral:
\begin{multline}
    \frac{1}{\hbar^2}\int \limits_0^{\infty} \dd\tau C_{3'4'}(\tau)e^{i \omega_{10} \tau}\\=\sum_{\nu}  \frac{\gamma_{\nu}}{2\pi}\int\limits_{0}^{\infty} \dd \tau \int\limits_{-\infty}^{\infty} \dd \omega \,\frac{1}{4} \frac{\Omega_R^2}{(\omega-\omega_{\nu})^2+\gamma^2_{\nu}/4}  e^{-\mathrm{i}(\omega-\omega_{10}) t}\\
    = \sum_{\nu}  \frac{\Omega^2_R}{2} \frac{\gamma_{\nu}/4}{(\omega_{10}-\omega_{\nu})^2+\gamma^2_{\nu}/4} \\ -i\sum_{\nu}\mathcal{P} \int\limits_{-\infty}^{\infty} \frac{\Omega^2_R}{2\pi}\frac{\dd \omega}{\omega-\omega_{10}}\frac{\gamma_{\nu}/4}{(\omega-\omega_{\nu})^2+\gamma^2_{\nu}/4}\\
    =\frac{\Gamma_{\downarrow}^r}{2}-\mathrm{i}\Delta_{\Gamma_{\downarrow}^r}
\end{multline}
where we have defined the resonant decay rate $\Gamma_{\downarrow}^r$ and shift $\Delta_{\Gamma_{\downarrow}^r}$ as:
\begin{subequations}
\begin{align}
    \frac{\Gamma_{\downarrow}^r}{2} &= \sum_{\nu}\frac{\Omega_R^2}{2} \frac{\gamma_{\nu}/4}{(\omega_{10}-\omega_{\nu})^2 + \gamma_{\nu}^2/4}, \\
    \Delta_{\Gamma_{\downarrow}^r} &=  \sum_{\nu}\mathcal{P} \int\limits_{-\infty}^{\infty} \frac{\Omega^2_R}{2\pi}\frac{\dd \omega}{\omega-\omega_{10}}\frac{\gamma_{\nu}/4}{(\omega-\omega_{\nu})^2+\gamma^2_{\nu}/4} .
 \end{align}
\end{subequations}

For the third bath the correlation values read:
\begin{align}
    &C_{56}(\tau)\approx 0, \\
    &C_{65}(\tau) \approx \int\limits_0^\infty \dd \omega'' \, \hbar^2 \mu^2(\omega'') e^{\mathrm{i}\omega'' \tau}. 
\end{align}
By evaluating the time--frequency integral with the coefficient $C_{65}(\tau)$ we obtain:
\begin{multline}
 \frac{1}{\hbar^2} \int\limits_0^{\infty} \dd \tau \, C_{65}(\tau)e^{-\mathrm{i}\omega_{10}\tau}\\= \frac{1}{\hbar^2}\int\limits_0^{\infty} \dd \tau \, \int\limits_0^\infty \dd \omega'' \, \hbar^2 \mu^2(\omega'') e^{\mathrm{i}\omega'' \tau}  e^{-\mathrm{i}\omega_{10}\tau}\\= \pi \mu^2(\omega_{10}) +\mathrm{i} \pv \int\limits_0^{\infty} \frac{\dd\omega''}{\omega''-\omega_{10}} \mu^2(\omega'').
\end{multline}
We define the laser pumping rate $\Gamma_{\uparrow}$ and the shift $\Delta_{\Gamma_{\uparrow}}$ as:
\begin{subequations}
\begin{align}
    \frac{\Gamma_{\uparrow}}{2} &= \pi \mu^2(\omega_{10}),\\
    \Delta_{\Gamma_{\uparrow}} &=   \pv \int\limits_0^{\infty} \frac{\dd\omega''}{\omega''-\omega_{10}} \mu^2(\omega''),
 \end{align}
\end{subequations}
with the coupling constant $\mu^2(\omega)$ being:

\begin{equation}
    \mu^2(\omega)=\frac{d_{01}^2 I(\omega)}{2c\varepsilon_0\hbar^2}.
\end{equation}
We find $\mu^2(\omega)$ in Appendix \ref{appendix:LaserDriving} where we derive the expression of the parameter $\Gamma_{\uparrow}$ related to the laser output intensity $I(\omega)$ and demonstrate that it does not depend on the frequency distribution of the light source in the broadband limit. Additionally, we find the frequency shift induced by the laser source $\Delta_{\Gamma_{\uparrow}}$.

\subsection{Resulting Master Equation}
Now we can write the master equation in shorter form retaining only the non-vanishing correlation functions:
\begin{widetext}
\begin{equation}\label{Eq:BigMasterEquation1_2}
\begin{aligned}
     \Dot{\rho}_S(t) = -\frac{\mathrm{i}}{\hbar} \comm{\Hh_S}{\rho_S(t)}
     &- 
     \frac{1}{\hbar^2}\int\limits_0^{\infty} \dd\tau 
     \bigg\{
      \sum_{\nu} \comm{\ahd_{\nu}}{\ah_{\nu} e^{\mathrm{i}\omega_{\nu} \tau}\rho_S(t)}C_{12}(\tau) +  \comm{\rho_S(t)\ahd_{\nu} e^{-\mathrm{i}\omega_{\nu} \tau}}{\ah_{\nu}}C_{12}(-\tau)
     \\[0.7em]
     &+
          \sum_i \comm{\shd_i}{\sh_i e^{\mathrm{i}\omega_{10}\tau}\rho_S(t)}\left[C_{34}(\tau)-C_{3'4'}(\tau)\right]
      +  \comm{\rho_S(t)\shd_i e^{-\mathrm{i}\omega_{10}\tau}}{\sh_i}\left[C_{34}(-\tau)-C_{3'4'}(-\tau)\right]
     \\[0.7em] &+
     \sum_{i} \comm{\rho_S(t)\sh_i  e^{\mathrm{i}\omega_{10}\tau}}{  \shd_i}C_{65}(-\tau) +  \comm{ \sh_i}{ \shd_i e^{-\mathrm{i}\omega_{10}\tau}\rho_S(t)}C_{65}(\tau) 
     \bigg\}.
\end{aligned}
\end{equation}
Using all the rate parameters defined above, the master equation reads:
\begin{equation}\label{Eq:BigMasterEquation1_3}
\begin{aligned}
     \Dot{\rho}_S(t) = &-\frac{\mathrm{i}}{\hbar} \comm{\Hh_S}{\rho_S(t)}
     - 
     \Bigg\{
     \sum_{\nu} \comm{\ahd_{\nu}}{\ah_{\nu} \rho_S(t)}\left[\frac{\kappa}{2}-\mathrm{i}\Delta_{\kappa} \right] +  \comm{\rho_S(t)\ahd_{\nu} }{\ah_{\nu}}\left[\frac{\kappa}{2}+\mathrm{i}\Delta_{\kappa} \right]
     \\[1em] 
     &+
     \sum_i \comm{\shd_i}{\sh_i \rho_S(t)}\left[\left[\frac{\Gamma^{\mathrm{tot}}_{\downarrow}}{2}-\mathrm{i}\Delta_{\Gamma^{\mathrm{tot}}_{\downarrow}}\right]-\left[\frac{\Gamma_{\downarrow}^r}{2}-\mathrm{i}\Delta_{\Gamma_{\downarrow}^r}\right]\right]
     +
     \comm{\rho_S(t)\shd_i }{\sh_i}\left[\left[\frac{\Gamma^{\mathrm{tot}}_{\downarrow}}{2}+\mathrm{i}\Delta_{\Gamma^{\mathrm{tot}}_{\downarrow}}\right]-\left[\frac{\Gamma_{\downarrow}^r}{2}+\mathrm{i}\Delta_{\Gamma_{\downarrow}^r}\right]\right]
     \\[1em] 
     &+
     \sum_{i} \comm{\rho_S(t)\sh_i  }{  \shd_i}\left[\frac{\Gamma_{\uparrow}}{2}-\mathrm{i}\Delta_{\Gamma_{\uparrow}}\right] +  \comm{ \sh_i}{ \shd_i \rho_S(t)}\left[\frac{\Gamma_{\uparrow}}{2}+\mathrm{i}\Delta_{\Gamma_{\uparrow}}\right]
     \Bigg\}.
\end{aligned}
\end{equation}
Note that for the last bath correlation the shifts $\Delta_{\Gamma_{\uparrow}}$ have the opposite sign as compared to $\Delta_{\Gamma^{\mathrm{tot}}_{\downarrow}}$,$\Delta_{\Gamma^r_{\downarrow}}$ and $\Delta_{\kappa}$.

After using the Lindblad superoperator definition $\mathfrak{L}[\hat{X}]\rho=\hat{X}^{\dagger}\hat{X}\rho +\rho \hat{X}^{\dagger}\hat{X} - 2\hat{X}\rho \hat{X}^{\dagger}$ for a generic operator $\hat{X}$ and collecting all Lamb shifts we obtain the master equation describing cavity and spontaneous decay and laser pump rate:
\begin{equation}\label{Eq:BigMasterEquation2}
\begin{aligned}
     \Dot{\rho}_S(t) &= -\frac{\mathrm{i}}{\hbar} \comm{\Hh_S}{\rho_S(t)}
     - \mathrm{i}\sum_{\nu,i}  \bigg\{ \Delta_{\kappa}\comm{\ahd_{\nu} \ah_{\nu}}{\rho_S(t)}+\Delta_{\Gamma_{\downarrow}}\comm{\shd_i \sh_i}{\rho_S(t)} - \Delta_{\Gamma_{\uparrow}}\comm{\sh_i \shd_i}{\rho_S(t)} \bigg\}\\ &-
     \bigg\{
     \sum_{\nu,i} \frac{\kappa}{2} \mathfrak{L}[\ah_{\nu}]\rho_S(t) +
     \frac{\Gamma_{\downarrow}}{2}\mathfrak{L}[\sh_i]\rho_S(t) +
     \frac{\Gamma_{\uparrow}}{2} \mathfrak{L}[\shd_i]\rho_S(t)
     \bigg\},
\end{aligned}
\end{equation}
\end{widetext}
where $\Gamma_{\downarrow}=\Gamma^{\mathrm{tot}}_{\downarrow}-\Gamma_{\downarrow}^r$ and $\Delta_{\Gamma_{\downarrow}}=\Delta_{\Gamma^{\mathrm{tot}}_{\downarrow}}-\Delta_{\Gamma_{\downarrow}^r}$.
Absorbing the Lamb shifts into a redefined system Hamiltonian the master equation reads:
\begin{multline}
        \Dot{\rho}_S(t) = -\frac{\mathrm{i}}{\hbar} \comm{\Hh_S}{\rho_S(t)}\\-
     \sum_{\nu,i} \bigg\{
     \frac{\kappa}{2} \mathfrak{L}[\ah_{\nu}]+
     \frac{\Gamma_{\downarrow}}{2}\mathfrak{L}[\sh_i] +
     \frac{\Gamma_{\uparrow}}{2} \mathfrak{L}[\shd_i]
     \bigg\}\rho_S(t).
\end{multline}
The new frequencies in the system Hamiltonian $\Hh_S$ coming from the Lamb shift we redefine to be $$\omega_{10}+\Delta_{\Gamma_{\downarrow}}+\Delta_{\Gamma_{\uparrow}}\to \omega_{10}$$ and $$\omega_{\nu}+\Delta_{\kappa}\to \omega_{\nu}.$$

\section{Nested Open Quantum Systems}\label{Section:Step2}
We have now derived the master equation for spontaneous and cavity decay and accounted for laser pumping. Now we treat the remaining system Hamiltonian as the total Hamiltonian of the subsystem $\Hh_S\equiv \Hh$. Also, we define Hamiltonians from step 1 to be the new system, bath and interaction Hamiltonians for the step 2:
\begin{gather}
    \Hh_{S1}+\Hh_{S2}\equiv \Hh_{S},\\
    \Hh_{SB4}\equiv \Hh_{I},\\
    \Hh_{B4}\equiv \Hh_{B}.
\end{gather}
If the coupling between rovibrational states,  electronic transitions and photons is strong, it is convenient to transform the Hamiltonian using the polaron transformation $\Hh \to \hat{U}^{\dagger}\Hh \hat{U}$, where the polaron operator $\hat{U}$ reads:
\begin{equation}
    \hat{U}=e^{\sum_i\sqrt{S}\sh_i^z(\bh_i-\bhd_i)}.
\end{equation}
The Hamiltonian after applying the polaron transformation can be written as:
\begin{eqnarray}
    \Hh = \sum_{\nu, i} &&\frac{1}{2}\hbar \omega_{10} \sh_i^z +
 \hbar \omega_{\nu} \ahd_{\nu}\ah_{\nu} +
 \hbar\Omega \bhd_i\bh_i \\&&+   
 \hbar \Omega_{\nu}[\ah_{\nu}\shd_i\Dh_i+\ahd_{\nu}\sh_i\Dhd_i],
\end{eqnarray}
with the interaction Hamiltonian being
\begin{equation}
    \Hh_I=\sum_{\nu, i}\hbar \Omega_{\nu}[\ah_{\nu}\shd_i\Dh_i+\ahd_{\nu}\sh_i\Dhd_i]
\end{equation}
and the displacement operator $\Dh=e^{2\sqrt{S}(\bhd_i-\bh_i)}$.

To obtain the master equation for $\Hh_I$, here we use a more general treatment, called projection operator technique  \cite{gardiner2004quantum}. Because now $\Hh_I$ in the interaction picture will not only evolve unitarily but will exhibit an exponential decay coming from $\kappa, \Gamma_{\downarrow}$ and $\Gamma_{\uparrow}$. We start with the master equation from the previous section:
\begin{equation}\label{Eq:BigMasterEquation3}
\begin{aligned}
     \Dot{\rho} = &-\frac{\mathrm{i}}{\hbar} \comm{\Hh}{\rho}
     \\& -
     \sum_{\nu,i} \bigg\{
     \frac{\kappa}{2} \mathfrak{L}[\ah_{\nu}] +
     \frac{\Gamma_{\downarrow}}{2}\mathfrak{L}[\sh_i] +
     \frac{\Gamma_{\uparrow}}{2} \mathfrak{L}[\shd_i]
     \bigg\}\rho.
\end{aligned}
\end{equation}
Note, that we have redefined the density matrix of the subsystem as $\rho_S\equiv\rho$.

Let us redefine commutators and Lindblad dissipators into Liouville superoperator form:
\begin{equation}\label{Eq:Superoperator}
    \Lc_{B,S,I}\,\rho = -\frac{\mathrm{i}}{\hbar} \comm{\Hh_{B,S,I}}{\rho},
\end{equation}
and
\begin{equation}
    \Dc \rho = -
     \sum_{\nu,i} \bigg\{
     \frac{\kappa}{2} \mathfrak{L}[\ah_{\nu}] +
     \frac{\Gamma_{\downarrow}}{2}\mathfrak{L}[\sh_i] +
     \frac{\Gamma_{\uparrow}}{2} \mathfrak{L}[\shd_i]
     \bigg\}\rho.
\end{equation}
The master equation then reads:
\begin{equation}
    \dot{\rho}=(\Lc_B+\Lc_S+\Lc_I+\Dc)\rho=\Lc\rho.
\end{equation}

The interest is to derive the master equation for the subsystem of interest, which can be achieved by projecting on the relevant part of the density matrix $\Pc\rho=\Tr_B[\rho]\otimes\rho_B=\rho_S\otimes\rho_B$. The irrelevant part reads as $\Qc\rho=(1-\Pc)\rho$. We follow the derivation along the lines of \cite{gardiner2004quantum} (ch. 5.1.2) and obtain
\begin{multline}\label{Eq:2ndMasterEq2}
    \dot{\rho}_S=(\mathcal{L}_S+\mathcal{D})\rho_S(t)\\-\frac{1}{\hbar^2}\text{Tr}_B[H_I, \int\limits_0^{\infty} \text{d} s \,e^{(\mathcal{L}_B+\mathcal{L}_S+\mathcal{D})s}[H_I, \rho_S(t-s)\otimes\rho_B]]
\end{multline}
Up to second order expansion in coupling $g$ from $\Hh_I$ it can be shown that Eq. \eqref{Eq:2ndMasterEq2} can be written as \cite{mari2012cooling}:
\begin{multline}\label{Eq:2ndMasterEq3}
   \dot{\rho}_S=(\mathcal{L}_S+\mathcal{D})\rho_S(t)\\-\frac{1}{\hbar^2}\text{Tr}_B[H_I, \int\limits_0^{\infty} \text{d} s \,[e^{(\mathcal{L}^{\dagger}_B+\mathcal{L}^{\dagger}_S+\mathcal{D}^{\dagger})s}(H_I), \rho_S(t)\otimes\rho_B]] 
\end{multline}
where $e^{(\mathcal{L}^{\dagger}_B+\mathcal{L}^{\dagger}_S+\mathcal{D}^{\dagger})s}$ is acting only on $\Hh_I$ and the adjoint Lindbladian superoperator for an arbitrary operator $\hat{A}$ is defined as:
\begin{equation}
    \mathfrak{L}^{\dagger}[\hat{X}]\hat{A}=\hat{X}^{\dagger}\hat{X}\hat{A} +\hat{A} \hat{X}^{\dagger}\hat{X} - 2\hat{X}^{\dagger} \hat{A} \hat{X}
\end{equation}
The interaction Hamiltonian now corresponds to the dissipative interaction picture. 

Calculating the Liouville superoperator for all interaction Hamiltonian operators, we obtain the following relations:
\begin{gather}
    \ah(t)=e^{(\mathcal{L}^{\dagger}_B+\mathcal{L}^{\dagger}_S+\mathcal{D}^{\dagger})t}\ah=e^{(\mathcal{L}^{\dagger}_S+\mathcal{D}^{\dagger})t}\ah=\ah e^{-\mathrm{i}\omega_{\nu} t -\frac{\kappa}{2}t},\\
    \ahd(t)=\ahd e^{\mathrm{i}\omega_{\nu} t -\frac{\kappa}{2}t},\\
    \sh(t)=\sh e^{-\mathrm{i}\omega_{10} t  -\frac{\Gamma_{\downarrow}+\Gamma_{\uparrow}}{2}t},\\
    \shd(t)=\shd e^{\mathrm{i}\omega_{10} t  -\frac{\Gamma_{\downarrow}+\Gamma_{\uparrow}}{2}t},\\
    \Dh(t)=e^{\mathrm{i}\sum_i \hbar\Omega\bhd_i\bh t}\Dh e^{-\mathrm{i}\sum_i \hbar\Omega\bhd_i\bh t}=e^{2\sqrt{S}(\bhd_i e^{\mathrm{i}\Omega t}-\bh_i e^{-\mathrm{i}\Omega t})},\\
    \Dhd(t)=e^{2\sqrt{S}(\bh_i e^{-\mathrm{i}\Omega t}-\bhd_i e^{\mathrm{i}\Omega t})},
\end{gather}
where the last two expressions we obtain by using Baker–Campbell–Hausdorff formula. We have used the following relations to obtain $\ah(t)$ and $\sh(t)$ and their conjugates:
\begin{gather}
    \mathfrak{L}^{\dagger}[\ah]\hat{1}=\mathfrak{L}^{\dagger}[\sh]\hat{1}=\mathfrak{L}^{\dagger}[\shd]\hat{1}=0,\\
    \mathfrak{L}^{\dagger}[\ah]\ah=\ah,\\
    \mathfrak{L}^{\dagger}[\ah]\ahd=\ahd,\\
    \mathfrak{L}^{\dagger}[\sh]\shd=\mathfrak{L}[\shd]\shd=\shd,\\
    \mathfrak{L}^{\dagger}[\sh]\sh=\mathfrak{L}[\shd]\sh=\sh.
\end{gather}

We are now in a position to write the master equation by evaluating the double commutator under the integral in Eq. \eqref{Eq:2ndMasterEq3}, which reads:
\begin{equation}\label{Eq:BigMasterEquation1_4}
\begin{aligned}
     \Dot{\rho}_S(t) = &-\frac{\mathrm{i}}{\hbar} \comm{\Hh_S}{\rho_S(t)}+\mathcal{D}\rho_S(t)
     \\ &- 
     \int\limits_0^{\infty} \dd\tau    
     \sum_{\nu,i} \bigg\{
     \comm{\ah_{\nu}  \shd_i}{\ahd_{\nu}  \sh_i  e^{\mathrm{i}\delta_{\nu} \tau}\rho_S(t)}C_{78}(-\tau)
     \\&+
     \comm{\rho_S(t)\ahd_{\nu}  \sh_i  e^{\mathrm{i}\delta_{\nu} \tau}}{\ah_{\nu}  \shd_i}C_{87}(\tau)
     \\ &+
     \comm{\ahd_{\nu}  \sh_i}{\ah_{\nu}  \shd_i e^{-\mathrm{i}\delta_{\nu} \tau}\rho_S(t)}C_{87}(-\tau) 
     \\&+
     \comm{\rho_S(t)\ah_{\nu}  \shd_i e^{-\mathrm{i}\delta_{\nu} \tau}}{\ahd_{\nu}  \sh_i}C_{78}(\tau)\bigg\},
\end{aligned}
\end{equation}
where $\delta_{\nu}=\omega_{\nu}-\omega_{10}$.

To cast this result into a simpler form, we define the quantities $K(\delta_{\nu}), K(-\delta_{\nu}),K^*(\delta_{\nu}), K^*(-\delta_{\nu})$:
\begin{multline}
    K(\delta_{\nu})=\int\limits_0^{\infty} \dd \tau \, C_{87}(-\tau)  e^{-\mathrm{i}\delta_{\nu} \tau}\\[-1em]=\hbar^2 \Omega_{\nu}^2\int\limits_0^{\infty} \dd \tau \, \langle\Dhd_i\Dh_i(\tau) \rangle  e^{-\mathrm{i}\delta_{\nu} \tau}e^{-\frac{\Gamma}{2} \tau},
\end{multline}
    \vspace{-8mm}
\begin{multline}
    K(-\delta_{\nu})=\int\limits_0^{\infty} \dd \tau \, C_{78}(-\tau)  e^{\mathrm{i}\delta_{\nu} \tau}\\[-1em]=\hbar^2 \Omega_{\nu}^2\int\limits_0^{\infty} \dd \tau \, \langle\Dhd_i\Dh_i(\tau) \rangle  e^{\mathrm{i}\delta_{\nu} \tau} e^{-\frac{\Gamma}{2} \tau},
\end{multline}
        \vspace{-8mm}
\begin{multline}
    K^*(\delta_{\nu})=\int\limits_0^{\infty} \dd \tau \, C_{87}(\tau)  e^{\mathrm{i}\delta_{\nu} \tau}\\[-1em]=\hbar^2 \Omega_{\nu}^2\int\limits_0^{\infty} \dd \tau \, \langle\Dhd_i\Dh_i(-\tau) \rangle  e^{\mathrm{i}\delta_{\nu} \tau} e^{-\frac{\Gamma}{2} \tau},
\end{multline}
        \vspace{-8mm}
\begin{multline}
    K^*(-\delta_{\nu})=\int\limits_0^{\infty} \dd \tau \, C_{78}(\tau)  e^{-\mathrm{i}\delta_{\nu} \tau}\\[-1em]=\hbar^2 \Omega_{\nu}^2\int\limits_0^{\infty} \dd \tau \, \langle\Dhd_i\Dh_i(-\tau) \rangle  e^{-\mathrm{i}\delta_{\nu} \tau} e^{-\frac{\Gamma}{2}\tau},
\end{multline}
with $\Gamma=\kappa+\Gamma_{\downarrow}+\Gamma_{\uparrow}$.
We have used the relation
\begin{equation}
   \langle \Dh_i(\tau) \Dhd_i \rangle = \langle \Dhd_i(\tau) \Dh_i \rangle,
\end{equation}
which follows from the relations $\Dhd(\alpha)=\Dh(-\alpha)$, and $\braket{n,-\alpha}{n, -\beta}=\braket{n,\alpha}{n, \beta}$. With these definitions, the master equation reads:
\begin{equation}\label{Eq:BigMasterEquation31}
\begin{aligned}
     \Dot{\rho}_S(t) = -\frac{\mathrm{i}}{\hbar} &\comm{\Hh_S}{\rho_S(t)}  +\mathcal{D}\rho_S(t) 
     \\ -
     \sum_{\nu,i}&\comm{\ah_{\nu}  \shd_i}{\ahd_{\nu}  \sh_i  \rho_S(t)}K(-\delta_{\nu}) \\ +& \comm{\rho_S(t)\ahd_{\nu}  \sh_i  }{\ah_{\nu}  \shd_i}K^*(\delta_{\nu})\\ +&  \comm{\ahd_{\nu}  \sh_i}{\ah_{\nu}  \shd_i \rho_S(t)}K(\delta_{\nu}) \\ +& \comm{\rho_S(t)\ah_{\nu}  \shd_i }{\ahd_{\nu}  \sh_i}K^*(-\delta_{\nu}).
\end{aligned}
\end{equation}
Separating the coefficients $K(\delta)$ into real and imaginary parts we can define constants for absorption and emission out of or into the cavity modes:
\begin{align}
    \Gamma(\pm\delta_{\nu})=2 K'(\pm\delta_{\nu}),
\end{align}
where $K'(\pm\delta_{\nu}) \equiv \Re K(\pm\delta_{\nu})$. With these definitions and collecting terms in Eq. \eqref{Eq:BigMasterEquation31} into Lindbladian forms, we obtain a final master equation that describes the dye--photon dynamics in a Photon-BEC setup:
\begin{eqnarray}\label{Eq:BigMasterEquation4}
     \Dot{\rho}_S = &&-\frac{\mathrm{i}}{\hbar} \comm{\Hh_S'}{\rho_S} - 
     \sum_{\nu,i} \bigg\{\frac{\Gamma_{\downarrow}}{2}\mathfrak{L}[\sh_i]\ +
     \frac{\kappa}{2} \mathfrak{L}[\ah_{\nu}]\nonumber  +
     \frac{\Gamma_{\uparrow}}{2}\mathfrak{L}[\shd_i]\\&& +
     \frac{\Gamma(\delta_{\nu})}{2}\mathfrak{L}[\ah_{\nu} \shd_i]+\frac{\Gamma(-\delta_{\nu})}{2}\mathfrak{L}[\ahd_{\nu} \sh_i]\bigg\}\rho_S.
\end{eqnarray}
The modified system Hamiltonian $\Hh'_S$ has absorbed the Lamb shifts and reads:
\begin{multline}
    \Hh'_S=\hbar\sum_{\nu, i} \Big[\omega_{10}+K''(-\delta_{\nu})\Big]\shd_i\sh_i + \Big[\omega_{\nu} +K''(\delta_{\nu})\Big]\ahd_{\nu}\ah_{\nu}\\+\Big[K''(-\delta_{\nu})-K''(\delta_{\nu})\Big]\ahd_{\nu}\ah_{\nu}\shd_i\sh_i, 
\end{multline}
where $K''(\pm \delta_{\nu})$ denotes the imaginary part of $K(\pm\delta_{\nu})$.

\section{Conclusions}
We have constructed a general theory and derived the necessary parameters to describe photon Bose--Einstein condensation in a dye-filled cavity using a microscopic description of the molecule--photon interaction. Adding to the dissipative previous studies, we have derived parameters $\Gamma_{\uparrow}, \Gamma_{\downarrow}$ and $\kappa$, where the last two depend on the geometry of the system. Also, we have demonstrated that all the rates (except $\Gamma_{\uparrow}$) are related to Green's tensor, which is essentially related to the geometrical setup. Furthermore, we have shown how $\Gamma(\pm\delta_{\nu})$ are influenced by $\Gamma_{\uparrow}, \Gamma_{\downarrow}$ and $\kappa$ since molecule rovibrational states are influenced by cavity and spontaneous decay and laser pumping. 

The next step will be to apply this technique to different geometries and calculate the threshold of the condensate. The simplest geometry for which Green's tensor can be analytically calculated is the planar cavity. For example, the spontaneous emission rate $\Gamma_{\downarrow}$ would then be calculated from the imaginary part of the Green's tensor multiplied by the dipole moment of the molecule, as demonstrated in the main section. If the cavity mirrors are highly reflective, it is relatively easy to calculate the cavity decay $\kappa$. It can be obtained from Green's tensor intrinsic structure by taking a Taylor expansion of the denominator in Green's tensor, which is responsible for multiple reflections from mirrors. For the laser pump rate $\Gamma_{\uparrow}$ one needs to know the laser intensity and dipole moment of the molecule.

For a more realistic setup, mirrors with a spherical curvature even including two dips \cite{kurtscheid2019thermally} can be considered, which are used in real experiments. Naturally, the complexity of calculating Green's tensor for such a system largely increases.

\begin{acknowledgments}
We want to thank Alessandra Colla, Andrea Mari, Andreas Ketterer, Axel U. J. Lode, Cyriaque Genet, David Steinbrecht, Dominik Lentrodt, Heinz-Peter Breuer, Peter Kirton, Yaroslav Gorbachev, Yue Ma for fruitful discussions.
The QUSTEC programme has received funding from the European Union's Horizon 2020 research and innovation programme under the Marie Skłodowska-Curie grant agreement number 847471. 
\end{acknowledgments}

\appendix

\section{Cavity Decay Demonstration}\label{appendix:CavityDecay}
Here we demonstrate that photons decay with the same rate as Lorentzian linewidth describing the quality of the cavity $\gamma$. We have defined the creation and annihilation operators in the cavity to be \cite{oppermann2018quantum}:
\begin{equation}\label{Eq1:BosonOperatorsCavity_appendix}
\begin{aligned}
    \ahd_{\nu}=\sqrt{\frac{\gamma_{\nu}}{2\pi}}\int\limits_{-\infty}^{\infty} \dd \omega \, \frac{\ahd(\omega)}{\omega-\omega_{\nu}-\mathrm{i}\gamma_{\nu}/2},\\
    \ah_{\nu}=\sqrt{\frac{\gamma_{\nu}}{2\pi}}\int\limits_{-\infty}^{\infty} \dd \omega \, \frac{\ah(\omega)}{\omega-\omega_{\nu}+\mathrm{i}\gamma_{\nu}/2}.
\end{aligned}
\end{equation}
As we will show, the effective dynamics of these non-monochromatic narrow-band operators will be non-unitary and decay with a rate $\gamma_{\nu}$ for mode $\nu$.
We start from Heisenberg equations of motion for $\ah(\omega)$ and their conjugate:
\begin{equation}\label{Eq1:HeisenbergEOM}
\begin{aligned}
    \ahddot(\omega)=\mathrm{i}\omega\ahd(\omega),\\
    \ahdot(\omega)=-\mathrm{i}\omega\ah(\omega).
\end{aligned}
\end{equation}
Taking the time derivative of Eqs. \eqref{Eq1:BosonOperatorsCavity_appendix} and using Eqs. \eqref{Eq1:HeisenbergEOM}, we obtain:
\begin{equation}\label{E1q:BosonOperatorsCavityEOM1}
\begin{aligned}
    \ahddot_{\nu}=\sqrt{\frac{\gamma_{\nu}}{2\pi}}\int\limits_{-\infty}^{\infty} \dd \omega \, \frac{i\omega\ahd(\omega)}{\omega-\omega_{\nu}-\mathrm{i}\gamma_{\nu}/2},\\
    \ahdot_{\nu}=\sqrt{\frac{\gamma_{\nu}}{2\pi}}\int\limits_{-\infty}^{\infty} \dd \omega \, \frac{-i\omega\ah(\omega)}{\omega-\omega_{\nu}+\mathrm{i}\gamma_{\nu}/2}.
\end{aligned}
\end{equation}
These equations can be rewritten in the form
\begin{equation}\label{Eq1:BosonOperatorsCavityEOM6}
\begin{aligned}
    \ahddot_{\nu}=\mathrm{i}\hat{F}^{\dagger} +(\mathrm{i}\omega_{\nu}-\frac{\gamma_{\nu}}{2})\ahd_{\nu},\\
    \ahdot_{\nu}=-\mathrm{i}\hat{F} +(-\mathrm{i}\omega_{\nu}-\frac{\gamma_{\nu}}{2})\ah_{\nu},
    \end{aligned}
\end{equation}
with $\hat{F}=\sqrt{\frac{\gamma_{\nu}}{2\pi}}\int \dd \omega \ah(\omega)$.
To find the equation of motion of the number operator $\expval{\ahd_{\nu}\ah_{\nu}}$, we use Eqs. \eqref{Eq1:BosonOperatorsCavityEOM6} and the chain rule. We start by taking derivative of the expectation value:
\begin{equation}
    \dv{t}\expval{\ahd_{\nu}\ah_{\nu}}= -\gamma_{\nu} \expval{\ahd_{\nu}\ah_{\nu}} -\mathrm{i}\expval{\ahd_{\nu}\hat{F}} + \mathrm{i}\expval{\hat{F}^{\dagger}\ah_{\nu}}.
\end{equation}
Let us evaluate the last two terms:
\begin{equation}
    -\mathrm{i}\expval{\ahd_{\nu}\hat{F}} 
    = 
    -\mathrm{i} \frac{\gamma_{\nu}}{2\pi}\int\limits_{-\infty}^{\infty} \dd \omega \, \frac{n(\omega)}{\omega-\omega_{\nu}-\mathrm{i}\gamma_{\nu}/2},
\end{equation}
\begin{equation}
    \mathrm{i}\expval{\hat{F}^{\dagger}\ah_{\nu}} 
    = 
    \mathrm{i} \frac{\gamma_{\nu}}{2\pi}\int\limits_{-\infty}^{\infty} \dd \omega \,   \frac{n(\omega)}{\omega-\omega_{\nu}+\mathrm{i}\gamma_{\nu}/2},
\end{equation}
where we have used the fact that $\expval{\ahd(\omega)\ah(\omega')}=n(\omega)\delta(\omega-\omega')$.
Summing these terms:
\begin{align}
   \mathrm{i}\expval{\hat{F}^{\dagger}\ah_{\nu}}-\mathrm{i}\expval{\ahd_{\nu}\hat{F}} 
   &= 
   \frac{\gamma^2_{\nu}}{2\pi}\int\limits_{-\infty}^{\infty} \dd \omega \,   \frac{ n(\omega)}{(\omega-\omega_{\nu})^2+(\gamma_{\nu}/2)^2},
\end{align}
the total time evolution equation reads:
\begin{multline}\label{Eq:PhotonDecay}
    \dv{t}\expval{\ahd_{\nu}\ah_{\nu}}= -\gamma_{\nu} \expval{\ahd_{\nu}\ah_{\nu}} \\+ \frac{\gamma^2_{\nu}}{2\pi}\int\limits_{-\infty}^{\infty} \dd \omega \,   \frac{ n(\omega)}{(\omega-\omega_{\nu})^2+(\gamma_{\nu}/2)^2},
\end{multline}
where $n(\omega)$ is thermal photon number following Bose--Einstein distribution:
\begin{equation}\label{eq:BoseEinsteinDistribution}
    n(\omega) = \frac{1}{e^{\frac{\hbar \omega}{k_B T}}-1}.
\end{equation}

In the experimental setup the thermal excitations are much smaller than the cavity excitations, thus, $n(\omega)\approx 0$. Eq. \eqref{Eq:PhotonDecay} then shows that the photon decay rate $\kappa$ is identical with the width of the resonance $\gamma_{\nu}$.

\section{Correlation Coefficients for the Total Photon Field Bath}\label{appendix:GreenTensor}

In this section we demonstrate explicitly how $C_{34}(\tau)$ and $C_{43}(\tau)$ are calculated. Before carrying out the calculation we mention that the average value of fundamental fields reads:
\begin{subequations}\label{Eq:BathAverages}
\begin{align}
     &\big\langle \fh(\rb, \omega)  \fhdp(\rb', \omega') \big\rangle = [n(\omega')+1]\vb{\boldsymbol{\delta}(r-r')} \delta_{\lambda \lambda'}\delta(\omega-\omega'),\\
          &\big\langle \fhd(\rb, \omega)  \fhp(\rb', \omega') \big\rangle = n(\omega')\vb{\boldsymbol{\delta}(r-r')} \delta_{\lambda \lambda'}\delta(\omega-\omega'),
     \end{align}
\end{subequations}
where $n(\omega)$ is the same as in Eq. \eqref{eq:BoseEinsteinDistribution}.

 Note that $C_{33} = C_{44} = 0$, because 
$
    \big\langle \fh(\rb, \omega)  \fhp(\rb', \omega') \big\rangle = \big\langle \fhd(\rb, \omega)  \fhdp(\rb', \omega') \big\rangle = 0.
$

The explicit calculation of the correlation coefficient $C_{34}(\tau)$ is as follows: we take the average of the product of the bath operators $\langle B_3 B_4(-\tau)\rangle$ from Eqs. \eqref{Eq:Bath2Operators} which is just an average over $\fh$ operators.

Simplifying the result by using the integral relation \cite{buhmann2013dispersion} \begin{multline}\label{Eq:IntegralRelation}
\sum\limits_{\lambda=e,m} \int \dd^3s   \,   \Gbl(\rb, \vb{s}, \omega) \cdot \Gbldp(\rb', \vb{s}, \omega') \\= \frac{\hbar\mu_0}{\pi} \omega^2 \Im \Gb(\rb, \rb', \omega),
\end{multline}
we find:
\begin{multline}
 C_{34}(\tau)\\=\int\limits_0^\infty \dd\omega \, \frac{\hbar\mu_0}{\pi} \omega^2 \vb{d}_{10} \cdot \Im\Gb(\rb_i, \rb_i, \omega)  \cdot \vb{d}_{01}[n(\omega)+1]e^{-\mathrm{i}\omega \tau}.
\end{multline}

The coefficient $C_{43}(\tau)$ is calculated in similar manner:
\begin{multline}\label{Eq:C43}
    C_{43}(\tau) = \langle B_4 B_3(-\tau) \rangle \\   = \int\limits_0^\infty \dd\omega \, \frac{\hbar\mu_0}{\pi} \omega^2 \vb{d}_{10} \cdot \Im\Gb(\rb_i, \rb_i, \omega)  \cdot \vb{d}_{01}n(\omega)e^{\mathrm{i}\omega \tau}.
\end{multline}

\section{Derivation of Laser Driving Constant}\label{appendix:LaserDriving}
In this section, we derive the laser driving constant from the properties of the laser. We follow a similar procedure as Loudon (ch. 2) \cite{loudon2000quantum}. We start from the Hamiltonian that describes the interaction of a two-level atom with an incoherent, broad-band classical light field:
\begin{multline}\label{Eq:LaserDrivingHamiltonian}
    \Hh = \Hh_A + \Hh_I = \frac{1}{2}\hbar\omega_{10} \sh_z \\+ \vb{\hat{d}}\cdot \int \limits_0^{\infty} \dd \omega \, \big[ \vb{E}(\rb_i, \omega) e^{-\mathrm{i}\omega t}e^{-\mathrm{i}\phi_{\omega}} +\vb{E}^*(\rb_i, \omega) e^{\mathrm{i}\omega t} e^{\mathrm{i}\phi_{\omega}}\big],
\end{multline}
where $\vb{E}(\rb_i, \omega)$ is the electric field at molecule's position $\rb_i$ at frequency $\omega$. The phase for each frequency is described by $\phi_{\omega}$ and $\vb{\hat{d}}=\vb{d}_{10}\dyad{1}{0}+\vb{d}_{01}\dyad{0}{1}$ is the dipole moment of the molecule.
To calculate time dynamics of excited atom state, we need to solve Schrödinger's equation $ \Hh \Psi= \mathrm{i}\hbar \dot{\Psi}$. We expand the wavefunction as linear superposition of orthonormal basis states which has a time dependence from atomic Hamiltonian:
\begin{equation}\label{Eq:WavefunctionExpansion}
    \ket{\Psi}=  C_0(t)e^{\mathrm{i}\frac{E_0}{\hbar}t}\ket{0} + C_1(t)e^{\mathrm{i}\frac{E_1}{\hbar}t}\ket{1},
\end{equation}
Transforming to the interaction picture, Schrödinger's equation reads:
\begin{equation}\label{Eq:InteractionSchrodingersEquation}
    \Hh_I \ket{\Psi}= \mathrm{i}\hbar (\dot{C}_0(t)\ket{0} + \dot{C}_1(t)\ket{1}).
\end{equation}
Multiplying Eq. \eqref{Eq:InteractionSchrodingersEquation} by $\bra{0}$ and $\bra{1}$ we obtain two differential equations:
\begin{equation}
\begin{aligned}
       \mathrm{i}\hbar \dot{C}_0(t)=&\vb{d}_{01}\cdot \int \limits_0^{\infty} \dd \omega \, \big[ \vb{E}(\rb_i, \omega) e^{-\mathrm{i}\omega t} e^{-\mathrm{i}\phi_{\omega}} \\&+\vb{E}^*(\rb_i, \omega) e^{i\omega t} e^{i\phi_{\omega}} \big] e^{-\mathrm{i}\omega_{10}t}C_1(t),  
       \\
       \mathrm{i} \hbar \dot{C}_1(t)=&\vb{d}_{10}\cdot \int \limits_0^{\infty}\dd \omega \, \big[ \vb{E}(\rb_i, \omega) e^{-\mathrm{i}\omega t} e^{-\mathrm{i}\phi_{\omega}} \\&+\vb{E}^*(\rb_i, \omega) e^{\mathrm{i}\omega t} e^{\mathrm{i}\phi_{\omega}} \big] e^{\mathrm{i}\omega_{10}t}C_0(t),  
\end{aligned}
\end{equation}
where we have used the fact that dipole operator has odd parity, meaning $\vb{d}_{00}=\vb{d}_{11}=0$. 
Using the rotating wave approximation and assuming that the dipole operator is real $ \vb{d}_{01}= \vb{d}_{10}$, the equations read:
\begin{equation}\label{Eq:TwoLevelSystemEq}
\begin{aligned}
       \mathrm{i}\hbar \dot{C}_0(t)=&\vb{d}_{01}\cdot \int \limits_0^{\infty} \dd \omega \,   \vb{E}^*(\rb_i, \omega) e^{\mathrm{i}(\omega-\omega_{10}) t} e^{\mathrm{i}\phi_{\omega}}  C_1(t),  \\
       \mathrm{i}\hbar \dot{C}_1(t)=&\vb{d}_{01}\cdot\int \limits_0^{\infty}\dd \omega \,    \vb{E}(\rb_i, \omega) e^{-\mathrm{i}(\omega-\omega_{10})t} e^{-\mathrm{i}\phi_{\omega}}  C_0(t).  
\end{aligned}
\end{equation}

To solve these equations we employ the perturbation expansion of $\vb{d}_{01}\cdot  \vb{E}(\rb_i, \omega)$ up to the first order, since $\vb{d}_{01}\cdot  \vb{E}(\rb_i, \omega)\ll\hbar\omega_{10}$. We pose the initial conditions, where the atom is assumed to be initially in the ground state, namely, $C_0(0)=1$ and $C_1(0)=0$. The solution for the first coefficient is constant $C_0(t)=1$. The solution for $C_1(t)$ reads:
\begin{equation}
   C_1(t)=\frac{\mathrm{i}}{\hbar}\vb{d}_{01}\cdot\int\limits_0^{\infty} \dd \omega \,  \vb{E}(\rb_i, \omega) \frac{1-e^{\mathrm{i}(\omega_{10}-\omega) t}}{\mathrm{i}(\omega_{10}-\omega)} e^{-\mathrm{i}\phi_{\omega}} . 
\end{equation}
Upon expressing the complex exponential in terms of the sine function we obtain:
\begin{multline}
   C_1(t)=-2\frac{\mathrm{i}}{\hbar}\vb{d}_{01}\cdot  \int\limits_0^{\infty} \dd \omega \,  \vb{E}(\rb_i, \omega)e^{\frac{\mathrm{i}}{2}(\omega_{10}-\omega)t} e^{-\mathrm{i}\phi_{\omega}}\\\times\frac{\sin{\left[\frac{1}{2}(\omega_{10}-\omega)t\right]}}{\omega_{10}-\omega}. 
\end{multline}
We are interested in the excited-state probability,
\begin{multline}
    \abs{C_1(t)}^2=\left(\frac{2d_{01}}{\hbar}\right)^2\\\times\abs{  \int\limits_0^{\infty} \dd \omega \,  \vb{E}(\rb_i, \omega)e^{\frac{\mathrm{i}}{2}(\omega_{10}-\omega)t}e^{-\mathrm{i}\phi_{\omega}}\frac{\sin{\left[\frac{1}{2}(\omega_{10}-\omega)t\right]}}{\omega_{10}-\omega}}^2,
\end{multline}
where for simplicity we have assumed that the dipole moment is parallel to the electromagnetic field $\vb{d}_{01}\parallel\vb{E}(\rb_i, \omega)$. For incoherent light, a phase average results in
\begin{equation}\label{Eq:PhaseAverageDiracDelta}
    \langle e^{\mathrm{i}(\phi_{\omega'}-\phi_{\omega})}\rangle=0, \,\,\,\mathrm{for} \,\,\,\omega\neq\omega'.
\end{equation}

Relating the electric field with the intensity of light using the well-known relation
\begin{equation}
    I = c\varepsilon_0 \vb{E}^2(\rb_i, t)=\int\limits_0^{\infty}d\omega\, I(\rb_i, \omega),
\end{equation}
the excitation probability can be expressed as:
\begin{equation}
    \abs{C_1(t)}^2 =\frac{2}{c\varepsilon_0\hbar^2} d^2_{01} \int\limits_0^{\infty} \dd \omega \, I(\rb_i, \omega)\frac{\sin^2{\left[\frac{1}{2}(\omega_{10}-\omega)t\right]}}{(\omega_{10}-\omega)^2}.  
\end{equation}

The intensity of a light source typically has some frequency distribution since it is not completely monochromatic. Thus, it can be described with a certain lineshape function $L(\omega)$ such that $I(\omega)=I_0 L(\omega)$ with $L(\omega_{10})=1$ and $I_0=I(\omega_{10})$. The excitation probability then reads:
\begin{equation}\label{Eq:ExcitProbab}
    \abs{C_1(t)}^2 =\frac{2}{c\varepsilon_0\hbar^2} d^2_{01} I_0\int\limits_0^{\infty} \dd \omega \, L(\omega) \frac{\sin^2{\left[\frac{1}{2}(\omega_{10}-\omega)t\right]}}{(\omega_{10}-\omega)^2}. 
\end{equation}

When we evaluate the integrals, we assume that arbitrary lineshape is centered at $\omega=0$ for the convenience of the calculations, which do not influence the physical results.
First we take $L(\omega)$ to be rectangular constant function centered at zero frequency, meaning that $L(\omega) = 1  $ in the interval $[-1/2\gamma, 1/2\gamma]$.  We calculate the following integral in the  limit $\gamma t \gg 1$:
\begin{equation}
  \int \limits_{-1/2\gamma}^{1/2\gamma} \dd \omega \,   \frac{\sin ^2\left(\frac{\omega t }{2}\right)}{\omega^2} \approx     \frac{\pi  t}{2} . 
\end{equation}
Secondly, we take $L(\omega)$ to be Gaussian centered at zero frequency $L(\omega)= e^{-\frac{1}{2}\frac{\omega ^2}{\gamma ^2}} $. This integral in the limit $\gamma t \gg 1$ is 
\begin{equation}
   \int \limits_{-\infty}^{\infty} \dd \omega \,  e^{-\frac{1}{2}\frac{\omega ^2}{\gamma ^2}} \frac{\sin ^2\left(\frac{\omega t }{2}\right)}{\omega^2}  \approx \frac{\pi  t}{2}.
\end{equation}
Thirdly, we evaluate same integral with a narrow Lorentzian profile centered at zero. This integral in the limit $\gamma t \gg 1$ is 
\begin{equation}
   \int \limits_{-\infty}^{\infty} \dd \omega \,  \frac{\frac{\gamma ^2}{4}}{ \left(\omega^2+\frac{\gamma ^2}{4}\right)} \frac{\sin ^2\left(\frac{\omega t }{2}\right)}{\omega^2} \approx   \frac{\pi  t}{2}.
\end{equation}

So regardless of the specific line shape, we find that in the limit $\gamma t \gg 1$
\begin{equation}
    \abs{C_1(t)}^2 \approx \frac{\pi d^2_{01}I_0}{c\varepsilon_0\hbar^2}  t=\Gamma_{\uparrow} t,
\end{equation}
where
\begin{equation}
    \Gamma_{\uparrow}=\frac{\pi d^2_{01}I_0}{c\varepsilon_0\hbar^2}.
\end{equation}

Using perturbation theory, we also calculate the frequency shift of the ground and excited state when subjected to laser light. By expanding the second equation in \eqref{Eq:TwoLevelSystemEq} up to zeroth order and solving it, we obtain the zeroth-order solution for $C_1^{(0)}(t)$:
\begin{multline}
    C_1^{(0)}(t)=\frac{\vb{d}_{01} }{\hbar}\cdot \int \limits_0^{\infty}\dd \omega \,  \vb{E}(\rb_i, \omega)  \frac{e^{-\mathrm{i}\phi_{\omega}}  e^{-\mathrm{i}(\omega-\omega_{10})t}}{\omega-\omega_{10}}C_0(t)\\
    -\frac{\vb{d}_{01} }{\hbar}\cdot\int \limits_0^{\infty}\dd \omega \,   \frac{ \vb{E}(\rb_i, \omega)  e^{-\mathrm{i}\phi_{\omega}}  }{\omega-\omega_{10}}C_0(t).
\end{multline}
We disregard the last term because it will have a oscillatory time dependence with frequency $\omega-\omega_{10}$. Substituting this into first equation of \eqref{Eq:TwoLevelSystemEq} we obtain the differential equation for $C_0(t)$:
\begin{multline}
    \mathrm{i} \dot{C}_0(t)= \frac{1}{\hbar^2}\int \limits_0^{\infty} \int \limits_0^{\infty}\dd \omega' \dd \omega \, \big[\vb{d}_{01}\cdot \vb{E}^*(\rb_i, \omega)\big]\big[\vb{d}_{01}\cdot \vb{E}(\rb_i, \omega')\big] \\\times\frac{ e^{\mathrm{i}(\omega-\omega') t} e^{\mathrm{i}(\phi_{\omega}-\phi_{\omega'})} }{\omega'-\omega_{10}}C_0(t).
\end{multline}
To simplify the double integral, we once more take a phase average and express the result in terms of light intensity to obtain
\begin{equation}\label{Eq:FrequencyShiftSolution}
    \mathrm{i} \dot{C}_0(t)=\frac{ d_{01}^2}{2 c \varepsilon_0 \hbar^2} \int \limits_0^{\infty}\dd \omega \,   \frac{I(\rb_i, \omega)}{\omega-\omega_{10}}C_0(t).
\end{equation}
This differential equation shows that the electric field induces a frequency shift (or light shift), which is proportional to the intensity and detuning of the laser field.

\providecommand{\noopsort}[1]{}\providecommand{\singleletter}[1]{#1}%

\end{document}